\documentclass[12pt, draftclsnofoot, onecolumn]{IEEEtran}
\IEEEoverridecommandlockouts
\usepackage{setspace}

\ifCLASSINFOpdf
\usepackage[pdftex]{graphicx}

\usepackage{verbatim}
\graphicspath{{figure/}}
\else
\fi
\ifCLASSOPTIONcaptionsoff
\usepackage[nomarkers]{endfloat}
\let\MYoriglatexcaption\caption
\renewcommand{\caption}[2][\relax]{\MYoriglatexcaption[#2]{#2}}
\fi
\usepackage{cite}
\usepackage{amsmath,amssymb,amsfonts}
\usepackage{cases}
\usepackage{graphicx}
\usepackage{epstopdf}
\usepackage{textcomp}
\usepackage{xcolor}
\def\BibTeX{{\rm B\kern-.05em{\sc i\kern-.025em b}\kern-.08em
	T\kern-.1667em\lower.7ex\hbox{E}\kern-.125emX}}
\usepackage{amsthm}
\usepackage{amssymb}
\usepackage{amsmath}

\newtheorem{proposition}{Proposition}

\usepackage{amsmath,amsfonts,amsthm,bm} % Math packages
\usepackage{booktabs}
\usepackage{float}
\makeatletter

\newcommand{\Rmnum}[1]{\expandafter\@slowromancap\romannumeral #1@}
\makeatother

\DeclareMathOperator{\minimize}{minimize}

\DeclareMathOperator{\subto}{subject \hspace{0.125em} to}
\usepackage{color, soul} 
\usepackage{subfigure}
\usepackage{threeparttable}
\usepackage{graphicx} % Required for the inclusion of images
\usepackage{graphics} % Required for the inclusion of images
\graphicspath{{figure/}}
\usepackage{amsmath} % Required for some math elements 
\usepackage{verbatim}%add comment
\usepackage{mathrsfs}
\usepackage{multirow}
\usepackage{float}
\usepackage{epstopdf}
\usepackage{mleftright}
\mleftright

\usepackage{tikz}

\usepackage[linesnumbered,ruled,vlined]{algorithm2e}
\usepackage{algpseudocode}
\SetKwInOut{Input}{\textbf{Input :}}  
\SetKwInOut{Output}{\textbf{Output :}} 
\usepackage{tikz}

\usepackage{array}
\renewcommand{\Re}{\operatorname{Re}}

\usepackage{subfigure}
\usepackage{color}
\definecolor{gray}{rgb}{0.6, 0.6, 0.6}
\definecolor{blue}{rgb}{0.0, 0.0, 1}

\usepackage{makecell}

\usepackage{enumerate}

\usepackage{xpatch}
\makeatletter
\def\changeBibColor#1{%
	\in@{#1}{intro-swipt-relay-8478360,intro-swipt-relay-9133290,intro-swipt-AF-7807356,intro-swipt-relay-8010881,  intro-zr-huang2019energy,concl-zr-6449245,intro-zr-8476597, intro-EHrelay-8653432, intro-EHrelay-8716549,intro-EHrelay-9107236, MC-8472907,book-anton1998calculus,EH-9069257,book1-gradshteyn2014table,book2-abramowitz1966handbook,ours-lu2020near}%  list of colored bib items
	\ifin@\color{black}\else\normalcolor\fi
}

\xpatchcmd\@bibitem
{\item}
{\changeBibColor{#1}\item}
{}{\fail}

\xpatchcmd\@lbibitem
{\item}
{\changeBibColor{#2}\item}
{}{\fail}
\makeatother
\begin{document}
\title{SER Analysis for SWIPT-Enabled Differential Decode-and-Forward Relay Networks}
\author{Yuxin~Lu,~\IEEEmembership{Student~Member,~IEEE,}		and~Wai~Ho~Mow,~\IEEEmembership{Senior~Member,~IEEE}
	\thanks{Yuxin Lu and Wai Ho Mow are with the Department of Electronic and Computer Engineering, the Hong Kong University of Science and Technology, Hong Kong S.A.R. (e-mail: ylubg@ust.hk; eewhmow@ust.hk).}
	\thanks{This work was supported by the Hong Kong Research Grants Council under GRF project no. 16233816.}
}

\maketitle

% As a general rule, do not put math, special symbols or citatioNS

%% Note that keywords are not normally used for peerreview papeRS.
%\begin{IEEEkeywords}
%	
%\end{IEEEkeywords}

% Note that keywords are not normally used for peerreview papeRS.
%\begin{IEEEkeywords}
%IEEEtran, journal, \LaTeX, paper, template.
%\end{IEEEkeywords}

% For peer review papeRS, you can put extra information on the cover
% \fi
%
% For peerreview papeRS, this IEEEtran command iNSerts a page break and
%	\IEEEpeerreviewmaketitle
%%--------------------------------------------------------------------------------------------------------
\begin{abstract}

In this paper, we analyze the symbol error rate (SER) performance of the simultaneous wireless information and power transfer (SWIPT) enabled three-node differential decode-and-forward (DDF) relay networks, which adopt the power splitting (PS) protocol at the relay. The use of non-coherent differential modulation eliminates the need for sending training symbols to estimate the instantaneous channel state information (CSI) at all network nodes, and therefore improves the power efficiency, as compared with the coherent modulation. However, performance analysis results are not yet available for the state-of-the-art detectors such as the maximum-likelihood detector (MLD) and approximate MLD. Existing works rely on the Monte-Carlo simulation method to show the existence of an optimal PS ratio that minimizes the overall SER. In this work, we propose a near-optimal detector with linear complexity with respect to the modulation size. We derive an  approximate SER expression and prove that the proposed detector achieves the full diversity order. Based on our expression, the optimal PS ratio can be accurately estimated without requiring any Monte-Carlo simulation. We also extend the proposed detector and its SER analysis for adopting the time switching (TS) protocol at the relay. Simulation results verify the effectiveness of our proposed detector and the accuracy of our SER results in various network scenarios for both PS and TS protocols.
    
\end{abstract}

\clearpage
\newpage
\section{Introduction}
The radio frequency (RF) signal has been widely used as the carrier for wireless information transmission (WIT). It has also become a new source for energy harvesting (EH) in the wireless power transfer (WPT) process \cite{ref0-zhou2013wireless}. In recent years, simultaneous wireless information and power transfer (SWIPT) has emerged as a promising technology to use the RF signal for WIT and WPT at the same time \cite{intro-tang2018energy}. SWIPT is an essential technology for various wireless systems to support different applications (see \cite{swipt-8114544,intro-zr-huang2019energy} and references therein), for example, for 5G communications to support the Internet of Things (IoT) applications \cite{intro-IoT-1-yang2017impact}. Most of the devices deployed in the IoT networks are of small sizes and low-powered, and harvesting energy from the RF signals can be a sustainable solution to provide them with convenient energy supplies \cite{intro-IoT-2-perera2018simultaneous}. 

In the SWIPT-enabled relay networks, the relay plays both the roles of EH for WPT and information processing for WIT. Two main receiver architectures are available for practical use at the relay \cite{intro-zhang2013mimo}, namely, the power splitting (PS) and time switching (TS) architectures. For the PS protocol, the received signal is separated in two portions which are used for the EH and information processing operations, respectively (see Section \ref{sec:sys-mod}). For the TS protocol, these two operations are performed in a time-division fashion, namely, the relay first performs EH and then performs information processing (see Section \ref{sec:sys-TS}). {\color{black} Many novel  contributions for the SWIPT-enabled relay networks have been proposed in the literature from the information-theoretic perspective. To maximize the network throughput, a hybrid EH protocol was developed in \cite{intro-ref6-atapattu2016optimal}, which combines the existing PS and TS protocols. The ergodic outage probability for the log-normal fading channels and  the generalized fading channels were studied in \cite{intro-swipt-relay-8010881} and \cite{intro-swipt-relay-8478360}, respectively, where accurate analytical expressions were derived.  
 For a two-way decode-and-forward (DF) relay network, the outage probability and the rate-energy  region were  characterized in \cite{intro-swipt-AF-7807356} and  \cite{intro-EHrelay-8653432}, respectively. The achievable rate and the  rate-energy trade-off were analyzed in \cite{intro-EHrelay-8716549} for a downlink amplify-and-forward (AF) relay network. To achieve secure communication, reference \cite{intro-EHrelay-9107236} studied  the secrecy outage probability, where three relay selection schemes were developed  for AF relaying. The energy accumulation method was considered in \cite{intro-swipt-relay-9133290} for adaptive relaying, where two novel schemes were developed and compared in terms of the  average throughput.}

One of the key challenges in the SWIPT-enabled relay networks is that the RF-powered relay nodes are energy-constrained \cite{intro-zr-8476597}, which restraints the use of high power consumption coding and decoding, modulation and demodulation techniques. Many existing works, such as \cite{ref0-zhou2013wireless,intro-swipt-relay-8010881,intro-swipt-relay-8478360,intro-swipt-AF-7807356,intro-EHrelay-8653432,intro-EHrelay-8716549,intro-ref6-atapattu2016optimal,intro-EHrelay-9107236} and \cite{intro-boshkovska2017secure,ref1-ye2018optimal}, use a coherent setup and assume that the instantaneous channel state information (CSI) is available at the receiving nodes. {\color{black}  }
 However, to acquire the instantaneous CSI requires frequent channel estimation, which causes additional power consumption and is clearly not friendly for such energy-constrained networks. Moreover, the additionally consumed power has negative impact on the future data relaying when the total power budget is fixed \cite{dm-df-ser-1-liu2015energy}. To address this issue, the power-efficient non-coherent differential modulation (DM) technique, which eliminates the channel estimation requirement, has become an attractive solution.  
% owing to its cost-effectiveness and energy-efficiency \cite{dm-af-1-mohjazi2018performance}.

%The absence of the channel estimation process reduces the hardware complexity, increases the energy efficiency, and therefore the non-coherent/differential modulation becomes an attractive solution to the energy-constrained SWIPT system \cite{dm-af-1-mohjazi2018performance}. 

Several works have been done to study the performance of the  SWIPT-enabled differential DF (DDF) and differential AF (DAF) relay networks in the literature (see \cite{dm-af-1-mohjazi2018performance, dm-af-ber-lou2017exact,dm-af-ser-1-liu2015noncoherent,dm-df-ser-1-liu2015energy,ours-lu2020near}). From the information-theoretic perspective, for DAF, some performance metrics were derived in \cite{dm-af-1-mohjazi2018performance}, such as the outage probability and the 
achievable throughput. Different from this perspective, in this paper, we aim to enhance the performance of the SWIPT-enabled relay networks from the symbol error rate (SER) perspective. In this case, for DAF, the SER performance of selection combining was studied in \cite{dm-af-ber-lou2017exact}, where analytical expressions were derived for the differential binary phase-shift keying (DBPSK) signal. The SER performances of the maximum-likelihood detectors (MLDs) based on the PS and TS protocols were analyzed in \cite{dm-af-ser-1-liu2015noncoherent} and \cite{dm-df-ser-1-liu2015energy}, respectively for DAF and DDF relay networks, and the approximate MLDs with lower complexities were also reported therein. In this paper, we focus on the DDF relay network and the state-of-the-art detectors in \cite{dm-df-ser-1-liu2015energy} serve as good performance benchmarks.  {\color{black} However, they still have high detection complexity.  The MLD and approximate MLD both involve multiple  Bessel function calculations and the MLD even involves multiple  integral calculations. Their detection complexities are quadratic and linear, respectively, with respect to the modulation size. }Moreover, their performance analysis results are not yet available in the literature, and the SERs of the respective systems were evaluated solely by the Monte-Carlo simulation method. This is possibly due to the non-closed-form detection metrics of the detectors. For example, the metrics adopted in \cite{dm-df-ser-1-liu2015energy} involve the modified Bessel function of the second kind and/or integral calculations.

To the best of our knowledge, for the SWIPT-enabled DDF relay networks, the SER analysis for an optimal or near-optimal detector has not been analyzed in the literature. {\color{black} Since the Monte Carlo method only gives numerical results, it cannot provide theoretical insights into the system performance, such as the achievable diversity order and the SER performance in terms of the network parameters, e.g., the PS ratio. Besides, it can be time-consuming \cite{MC-8472907}.}
Motivated by the above, we propose to address the low-complexity near-optimal detection problem and analyze the SER performance for the DDF single-relay network, using the differential phase-shift keying (DPSK) signals. {\color{black} We first present the results adopting the PS protocol at the relay and then extend our results for adopting the TS protocol.}
%adopting the PS protocol at the relay. % , without the need for performing Monte-Carlo simulations. % The Monte-Carlo simulation results in \cite{dm-df-ser-1-liu2015energy} show that there is an optimal value of the PS ratio that minimizes the SER.
%{\color{black} As a main advantage, The proposed approximate SER has an explicit expression, which can be used to analyze the SER performance with respects to various parameters. Besides, the proposed methods use arithmetic operations to estimate the optimized PS ratio, which is very efficient. By contrast, the existing Monte Carlo method only gives numerical results and can be time-consuming  due to the need of a large sample size \cite{MC-8472907}.}

{\color{black}
The main contributions are summarized as follows.
\begin{itemize}
	\item We propose a near-optimal detector with a closed-form metric, which performs very close to the state-of-the-art detectors in  \cite{dm-df-ser-1-liu2015energy} for various modulation sizes. It only involves  low-complexity arithmetic operations and has linear complexity with respect to the modulation size. By contrast, the existing detectors involve high-complexity Bessel function and/or integral calculations.
	\item For the SER analysis, we develop an approximate SER expression and prove that the proposed detector achieves the full diversity order of two.
	Simulation results verify its accuracy in various network scenarios. This theoretical result is novel in that the existing work \cite{dm-df-ser-1-liu2015energy} only showed the SER performance via Monte Carlo simulation.
	\item Our SER expression can be used to optimize the SER performance in terms of the network parameters such as the PS ratio.  Based on this expression, we propose two methods to  estimate the optimal value of the PS ratio that minimizes the SER. Both methods are verified by simulation to be very accurate.
	\item We also extend the proposed detector and the SER analysis results for adopting the TS protocol at the relay. An approximate SER expression is derived, based on which we prove that the proposed detector achieves the full diversity order of two. Simulation results verify the effectiveness of our detector and the accuracy of our SER results in various network scenarios, e.g., different channel conditions, modulation sizes, and PS/TS ratios.
\end{itemize}
}

The rest of the paper is organized as follows. Section \ref{sec:sys-mod} presents the system model. The proposed  detector and SER expression are derived in Section \ref{sec:det-bound}, {\color{black} followed by the diversity order analysis}. Section \ref{sec:ana} analyzes this SER expression and derives the optimal PS ratio. {\color{black} The extended results for the TS protocol is presented in Section \ref{sec:TS}. } Section \ref{sec:simu} presents simulation
results followed by conclusions in Section \ref{sec:con}. Related proofs are provided in the Appendix.

\textit{Notation:} $\Pr[X]$ denotes the probability of an event $X$. $\mathbb{E}[X]$ represents the expected value of $X$. $\Re \{x\}$ denotes the real part of a complex number $x$. $Q(x) \triangleq \frac{1}{\sqrt{2 \pi}} \int_{x}^{\infty} \exp (-z^2/2) dz$.
%%--------------------------------------------------------------------------------------------------------
\section{System Model}
\label{sec:sys-mod}
\begin{figure}[t]
	\centering
	\includegraphics[width=1\textwidth]{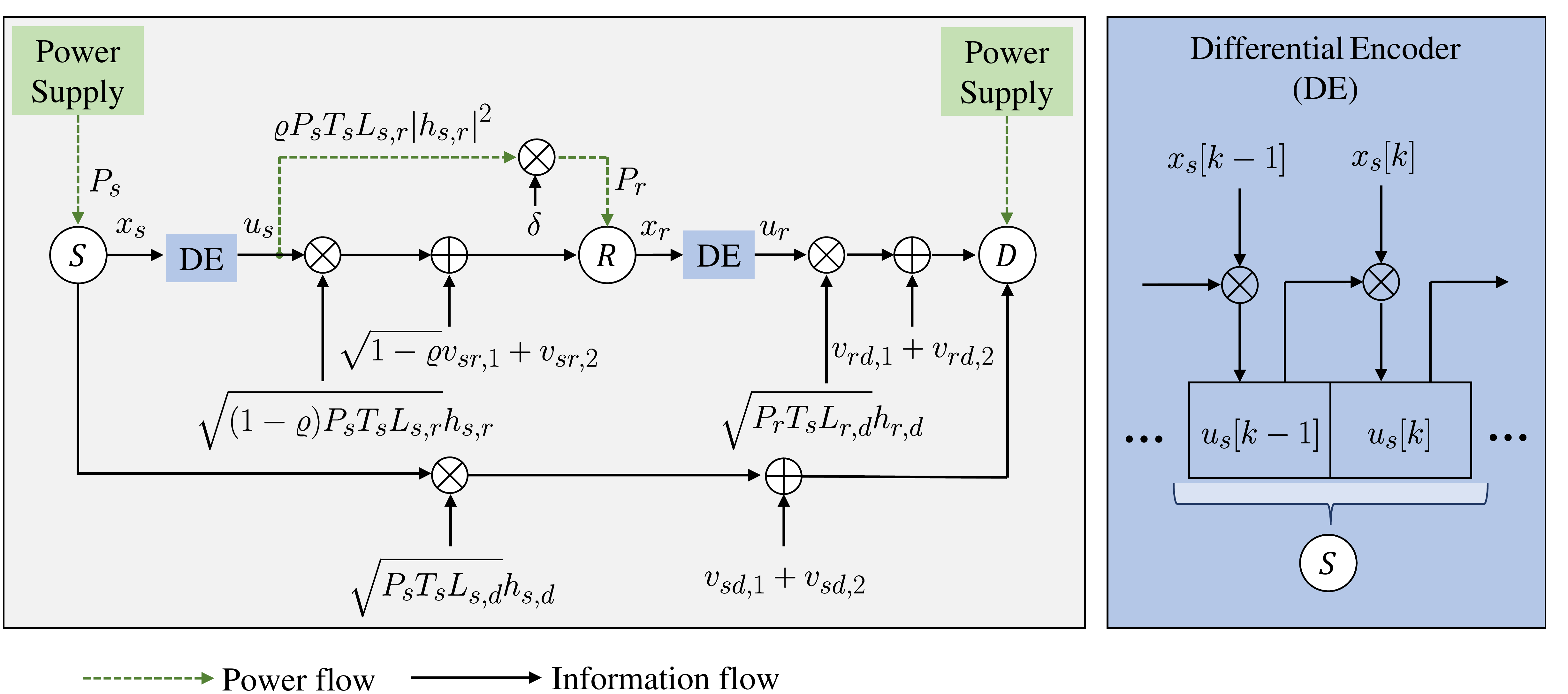}%{PS-channel-model}
	\caption{The system model of the $3$-node SWIPT-enabled PS-based DDF relay network, where the solid and dashed arrow lines denote the information and the power flows, respectively. \color{black} The source symbol $x_s$ is differentially encoded as $u_s$ for transmission, and similarly for the relay symbol $x_r$.}
	\label{fig:channel-model-PS}
\end{figure} %  
We consider a three-node SWIPT-enabled PS-based DDF relay network with one source ($S$), one half-duplex relay ($R$) and one destination ($D$). The system model is shown in Fig.~\ref{fig:channel-model-PS}, where the information and power flows are shown using the solid and dashed arrow lines, respectively. Assume $R$ has no CSI, while $D$ has only the statistical CSI of the $S-R$ link but no CSI of the other links. $S$ and $D$ have dedicated energy sources such as a battery or power grid, and $S$ transmits its message with a constant power $P_s$. However, $R$ has no power supply and can only harvest energy from the received signals from $S$ for detection and transmission. For the $I-J$ link, $(I,J) \in \{(s,r), (s,d), (r,d)\}$, assume small-scale Rayleigh fading $h_{I,J}$ and large-scale path loss $L_{I,J}$. Let $v_{IJ,1} \sim \mathcal{CN}(0, N_{IJ,1})$ denote the complex additive
white Gaussian noise (AWGN) at the receive antenna and $v_{IJ,2} \sim \mathcal{CN}(0, N_{IJ,2})$ denote the complex AWGN due to the circuit.

One symbol transmission consists of two time slots each with duration $T_s = T/2$. In the first time slot, $S$ transmits its signal to $R$ and $D$. In the second time slot, $R$ transmits its detected symbol to $D$ while $S$ remains silent. DPSK is used. Let $S$ select a symbol $x_s$ from the $M$-PSK alphabet, defined as $\mathcal{X} \triangleq  \{x_m = e^{j2 \pi (m-1)/M}, m=1,2, \dots ,M\}$, with equal probability. {\color{black} The differential encoder (DE) is shown in Fig.~\ref{fig:channel-model-PS}}, where $x_s$ is differentially encoded as $u_s$ for transmission, and $R$ uses the same approach to encode $x_r$ as $u_r$ for transmission. For the $k$-th symbol, we can write the DE  process as 
{\color{black}
\begin{align}
	u_I[k] =  u_I[k-1] x_I[k], \ k=1,2,3,\dots,
\end{align} where the initialization symbol is defined as $u_I[0]=1$ and $I \in \{s,r\}$ denotes the transmitting node (source or relay). 
}
\begin{figure}[t]
	\centering
	\includegraphics[width=0.6\textwidth]{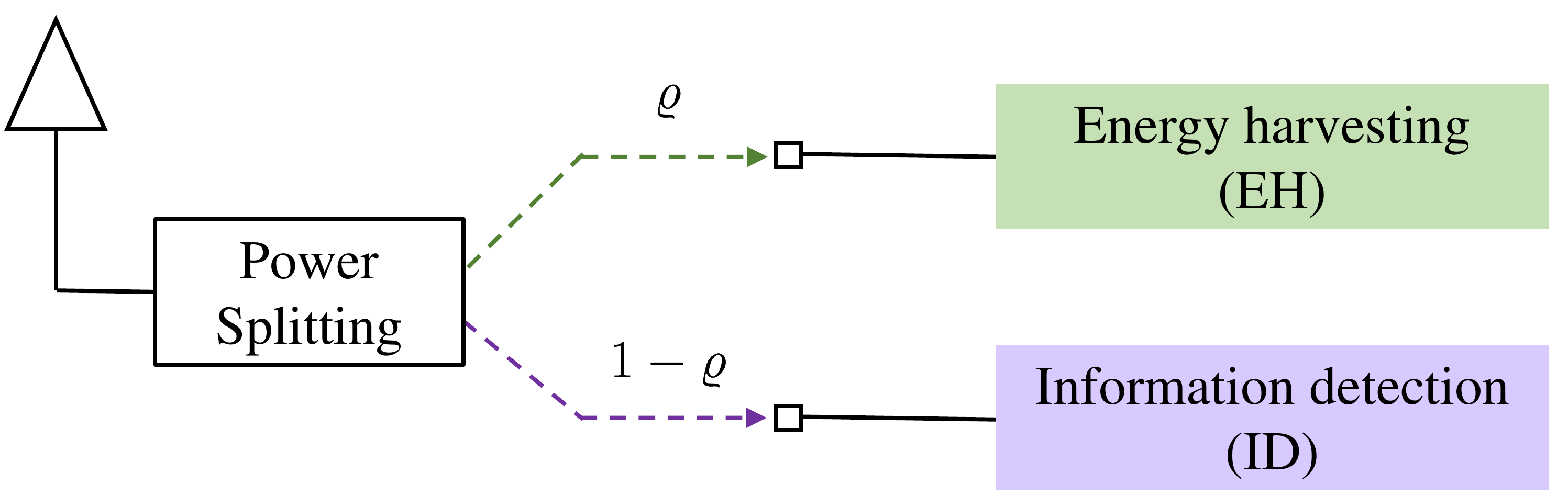}%{PS-channel-model}
	\caption{\color{black}Block diagram of the relay receiver architecture adopting  the PS protocol.}
	\label{fig:PS}
\end{figure} 

{\color{black}As shown in Fig.~\ref{fig:PS},} the PS protocol is adopted at $R$ for EH with the PS ratio $0 < \varrho < 1$. Specifically, $R$ splits its received signal from $S$ into two portions, a $\varrho$ portion for EH and a $1-\varrho$ portion for information detection (ID). We use the linear EH model here for simplicity, and leave the study of the non-linear model (such as that described in \cite{eh-non-li-boshkovska2015practical}) as future work. The harvested energy at $R$ is $E_r = \delta \varrho P_s L_{s,r} |h_{s,r}|^2 T_s$, and the transmission power at $R$ is
\begin{align}
	P_r = P_r(\varrho, h_{s,r}) = \frac{E_r}{T_s} = \delta \varrho P_s L_{s,r} |h_{s,r}|^2, \label{eq:Pr}
\end{align}
where $\delta \in (0,1]$ denotes the power conversion efficiency at $R$. The impact of the quality of the $S-R$ link on the transmission of the $R-D$ link is obvious since $P_r$ is directly affected by the term $L_{s,r} |h_{s,r}|^2$, which is a measure of the $S-R$ link quality. The received signal at $R$ for ID is
\begin{align}
y_{s,r}[k] = & \sqrt{(1-\varrho)P_s T_s L_{s,r}} h_{s,r} u_s[k] +  \sqrt{1-\varrho}v_{sr,1}[k] + v_{sr,2}[k] , \label{eq:ysr}
\end{align}
and the received signals at $D$ are
\begin{align}
y_{I,d}[k] = & \sqrt{P_I T_s L_{I,d}} h_{I,d} u_I[k] +  v_{Id,1}[k] + v_{Id,2}[k],  \label{eq:yrd}
\end{align}
where $I \in \{s,r\}$ denotes the transmitting node (source or relay).

For DM, by assuming the channel coefficients remain unchanged for at least two consecutive symbol intervals, the received signals in \eqref{eq:ysr} and \eqref{eq:yrd}  can also be written as
\begin{align} \label{eq:p2p}
y_{I,J}[k] = & y_{I,J}[k-1] x_I[k] + n_{I,J},
\end{align}
where $I \in \{s,r\}$ denotes the transmitting node, $J \in \{r,d\}$ denotes the corresponding receiving node, 
%\begin{align}
%	n_{s,r} = & \sqrt{1-\varrho}v_{sr,1}[k] + v_{sr,2}[k] - \notag \\ & x_s[k](\sqrt{1-\varrho}v_{sr,1}[k-1] + v_{sr,2}[k-1]), \\
%	n_{I,d} = & v_{Id,1}[k] + v_{Id,2}[k]- \notag \\ & x_I[k](v_{Id,1}[k-1] + v_{Id,2}[k-1])  ,
%\end{align}
$
n_{s,r} =  \sqrt{1-\varrho}v_{sr,1}[k] + v_{sr,2}[k] -  x_s[k](\sqrt{1-\varrho}v_{sr,1}[k-1] + v_{sr,2}[k-1]) \sim  \mathcal{CN} (0, 2(1-\varrho) N_{sr,1} + 2 N_{sr,2}) $, and $n_{I,d} =  v_{Id,1}[k] + v_{Id,2}[k]-  x_I[k](v_{Id,1}[k-1] + v_{Id,2}[k-1]) \sim  \mathcal{CN} (0, 2 N_{Id,1} + 2 N_{Id,2}) 
$. 

Note that for the detection at $R$, we adopt \cite[eq. (27)]{dm-df-ser-1-liu2015energy}, which is performed based on the relation of the two consecutively received symbols as shown in \eqref{eq:p2p}, and thus requires no CSI of the $S-R$ link. {\color{black} The statistical CSI associated with the $S-R$ link for ID at R is denoted as $\bar{\gamma}_{s,r}^{ID} (\varrho) = \frac{(1-\varrho) T_s  P_s L_{s,r} }{(1-\varrho) N_{sr,1} + N_{sr,2}}$ (calculated from \eqref{eq:ysr}), which is used at $D$ to calculate the average SER of ID at $R$, denoted as $\epsilon$, given by (c.f.  \cite[eq. (33)]{dm-df-ser-1-liu2015energy})
\begin{align} \label{eq:SER-p2p-DM-2}
\epsilon = \epsilon (\varrho)  \triangleq  
\begin{cases} 
\frac{ 1 }{2 \left[1+ \bar{\gamma}_{s,r}^{ID} (\varrho) \right]}, & M = 2,\\
1.03 \sqrt{\frac{1+\cos \frac{\pi}{M}}{2 \cos \frac{\pi}{M}}}   \left[ 1-
\sqrt{\frac{ (1-\cos \frac{\pi}{M})  \bar{\gamma}_{s,r}^{ID} (\varrho) }{1+  (1-\cos \frac{\pi}{M}) \bar{\gamma}_{s,r}^{ID} (\varrho)  }} \
\right], & M > 2.
\end{cases}
\end{align}
Note that $\epsilon $ will be used in the next section for developing our detector.
}
%%--------------------------------------------------------------------------------------------------------
\section{Proposed Detector and SER Analysis}
\label{sec:det-bound}
In this section, we introduce our linear-complexity detector, derive an SER expression for this detector, and analyze the diversity order. 
%%-------------------------------------------------
\subsection{Proposed Detector at the Destination}
\label{sec:det}
The optimal MLD for DM should find the source symbol that maximizes the conditional joint probability density of the received signals as
\begin{align}
	\max_{x_s \in \mathcal{X}} f(y_{s,d}[k]|x_s, y_{s,d}[k-1]) \sum_{x_r \in \mathcal{X}} \Pr(x_r|x_s) f(y_{r,d}[k]|x_r,y_{r,d}[k-1]).
\end{align} 
By using the average SER  of ID at $R$, i.e., $\epsilon$ in \eqref{eq:SER-p2p-DM-2}, to estimate the transition probability terms as {\color{black}$\Pr(x_r = x_s|x_s) = 1-\epsilon $ and $\Pr(x_r \neq x_s|x_s) = \epsilon/(M-1)$}, and applying the widely used max-sum approximation, which gives the excellent performance, especially at high signal-to-noise ratio (SNR) (c.f. \cite{qian2017near,kim2015low}), we obtain a near-optimal detection metric as 
\begin{align}
\max_{x_s \in \mathcal{X}} &   \{ f(y_{s,d}[k]|x_s, y_{s,d}[k-1]) \max \{ 
 (1-\epsilon)  f(y_{r,d}[k]|x_s,y_{r,d}[k-1]) 
, \notag \\
& \epsilon/(M-1) \max_{ \substack{x_r \in \mathcal{X}, x_r \neq x_s} } f(y_{r,d}[k]|x_r,y_{r,d}[k-1])  \} \} .  \label{eq:max-log}
\end{align}
Further, based on the observation that $\epsilon < 0.5$ is sufficient to ensure $1-\epsilon > \epsilon/ (M-1)$, we remove the constraint $x_r \neq x_s$ in \eqref{eq:max-log} and finally develop the proposed near-optimal detector, which is performed based on 
\begin{align}		
\hat{x}_s  
 = &	\arg  \max_{x_s \in \mathcal{X}}  \{ f(y_{s,d}[k]|x_s, y_{s,d}[k-1]) \max \{ 
	(1-\epsilon)f(y_{r,d}[k]|x_s,y_{r,d}[k-1]) 
	 , \notag \\ & \epsilon/(M-1)  \max_{ \substack{x_r \in \mathcal{X}, x_r \neq x_s} } f(y_{r,d}[k]|x_r,y_{r,d}[k-1])  \} \} % \label{eq:max-ML-metric-1}
	  \\
= &  \arg  \min_{x_s \in \mathcal{X}}  \Bigg\{ \frac{\Re \{ y_{s,d}^*[k] y_{s,d}[k-1] x_s\}}{N_{sd,1}+ N_{sd,2}}
+  \min \Bigg\{ \frac{\Re \{ y_{r,d}^*[k] y_{r,d}[k-1] x_s\}}{N_{rd,1} + N_{rd,2}}
+\eta,  \notag \\
&  \min_{x_r \in \mathcal{X}} \frac{\Re \{ y_{r,d}^*[k] y_{r,d}[k-1] x_r\}}{N_{rd,1} + N_{rd,2}} 
\Bigg\}
\Bigg\} ,  \label{eq:det}
\end{align}
where {\color{black}$ y_{r,d}$ is a function of $\varrho$ and $h_{s,r}$ since $P_r$ is a function of them (c.f. \eqref{eq:Pr} and \eqref{eq:yrd}), and $\eta = \eta(\varrho)  \triangleq \ln\frac{[1-\epsilon (\varrho)](M-1)}{\epsilon (\varrho)} $. Note that $\epsilon $ and $\eta$ are functions of $\varrho$ (c.f. \eqref{eq:SER-p2p-DM-2})} The complexity of this detector is linear with respect to the modulation size $|\mathcal{X}|=M$, because the enumerations over $x_s \in \mathcal{X}$ and $x_r \in \mathcal{X}$ are decoupled. 
%%-------------------------%%
{ \color{black}
	\subsection{ Detection Complexity Comparison}
	The MLD \cite{dm-df-ser-1-liu2015energy} involves integral calculations, which do not have closed-form solutions. To evaluate its complexity,  we adopt the Riemann sum method \cite{book-anton1998calculus}  to approximate the integral using a finite sum as 
\begin{align}
\int_{a}^{b} g(s) d s \approx  \sum_{n=1}^{S} g(s_n)  \frac{b-a}{S}, 
\end{align}
where the interval $[a,b]$ is divided into $S$ equal-length subintervals and $s_n$ denotes a point in the $n$-th subinterval that can be chosen. 
	\begin{table}[t] 
	\centering
	{ \color{black} 
		\caption{Number of operations per symbol detection required by MLD \cite{dm-df-ser-1-liu2015energy}, approximate MLD \cite{dm-df-ser-1-liu2015energy}, and the proposed detector for the  SWIPT-enabled DDF relay network using $M$-DPSK }
		\label{table:complexity}
		\begin{center}
			\noindent
			\begin{minipage}{1\textwidth}			\centering
				\begin{small}
					\begin{tabular}{c|cccc}
						
						\toprule
						\diaghead{\theadfont Diag ColumnmnHead}%
						{\textbf{Detector}}{\textbf{Operation}}&
						\thead{\textbf{Addition}}&\thead{
							\textbf{Multiplication}} &\thead{\textbf{Bessel function}} &\thead{\textbf{Table look-up}\footnote{\color{black}Exponential, logarithm, and trigonometric functions and their inverses are included.}} \\ 
						\midrule
						MLD & $M(7MS+7M)$ & $M(15MS+20M+8)$ &  $M^2$ & $M(4MS+M+1)$ \\ %
						Approximate MLD & $41M$ & $78M$ &  $4M$  & $13M$ \\	
						Proposed detector & $8M$ & $14M$ &  $0$ &  $1$ \\
						\bottomrule
					\end{tabular}
				\end{small}
			\end{minipage}
		\end{center} 
	}
\end{table}
As such, the integral calculation is converted to countable arithmetic operations.  
Table \ref{table:complexity} shows the number of operations required by the MLD, approximate MLD, and the proposed detector for detecting one $M$-DPSK symbol. We can see that the MLD and approximate MLD have $\mathcal{O}(M^2)$ and $\mathcal{O}(M)$  complexities, respectively, while the proposed detector has $\mathcal{O}(M)$ complexity. Besides, both the MLD and approximate MLD involve Bessel function calculations, and the numbers of calculations are $\mathcal{O}(M^2)$ and $\mathcal{O}(M)$, respectively. By contrast, the proposed detector has a closed-form metric and only involves low-complexity arithmetic operations.
}	
	
%%--------------------------------------------------------
\subsection{Approximate SER for the Detector}
% Other cases can be studied through adjusting the path loss $L_{I,J}$. 
Consider the case  $N_{IJ,1}=N_{IJ,2}=N_0/2$ for all links hereafter for simplicity,  the instantaneous SNR of the $I-J$ link is defined as $\gamma_{I,J} \triangleq \frac{P_s |h_{I,J}|^2}{N_0}$, and the corresponding average SNR is defined as $\bar{\gamma}_{I,J} \triangleq \frac{P_s }{N_0} \mathbb{E} [|h_{I,J}|^2]  $. Denote the real Gaussian random variables associated with the detection metric of the $I-d$ link, $I \in \{s,r\}$, as
\begin{align}
\omega_{I,d} (z_1, z_2) 
= &  \Re \{ y_{I,d}^*[k] y_{I,d}[k-1] (z_2-z_1)\} /N_0 \notag \\
\sim & \mathcal{N} (u_{I,d}(z_1, z_2), W_{I,d} (z_1, z_2)), \label{eq:omega}
\end{align}
and we have
\begin{align}
u_{I,d}(z_1, z_2) = & T_s L_{I,d} \Re \{
x_{I}^*(z_2-z_1)
\} \frac{P_I |h_{I,d}|^2}{N_0} , \label{eq:w-mean}\\
W_{I,d}(z_1, z_2) \approx  & T_s L_{I,d} |z_2-z_1|^2 \frac{P_I |h_{I,d}|^2}{N_0}. \label{eq:w-var}
\end{align}
The mean in \eqref{eq:w-mean} can be obtained by substitutions. For the approximate variance in \eqref{eq:w-var}, the approximation is due to
\begin{align}
|y_{I,d}[k]|^2  = & |\sqrt{P_I T_s L_{I,d}} h_{I,d} u_I[k] + n_{I,d}[k] + v_{I,d}[k]|^2 \notag \\
\approx & P_I T_s L_{I,d} |h_{I,d}|^2 , \label{eq:approx}
\end{align}
where high order noise terms are ignored \cite{ref1-bhatnagar2012decode}. % Note that for DPSK signals, there is $W_{I,d}(x_I, z_2) = -2u_{I,d}(x_I, z_2)$.

Using the defined variables in \eqref{eq:omega}, an approximate SER expression is derived in Appendix \ref{proof-prop-1}, and presented in \textbf{Proposition \ref{prop: 1}}, where $\check{\gamma} \triangleq [\gamma_{s,d}, \gamma_{r,d},\gamma_{s,r}]$, $g_{s,d} \triangleq  \sin^2  \left(\frac{\pi}{M}\right) T_s L_{s,d} $ and $ g_{r,d} \triangleq  \sin^2 \left(\frac{\pi}{M}\right)   T_s  L_{s,r} L_{r,d}$. $\mathcal{P}_C (\check{\gamma})$ and $\mathcal{P}_E (\check{\gamma})$ characterize the conditional SER performances for the two scenarios where the relay detects correctly and wrongly, respectively.
\begin{proposition}\label{prop: 1}
	For the SWIPT-enabled DDF single-relay network adopting the PS protocol, the overall SER of the proposed near-optimal detector is approximately equal to $  \mathcal{P}_C (\check{\gamma}) + \mathcal{P}_E  (  \check{\gamma} )$ for $M>2$, and $ \frac{1}{2} \mathcal{P}_C (\check{\gamma}) + \frac{1}{2} \mathcal{P}_E  (  \check{\gamma} )$ for $M=2$, where
	\begin{align}
	\mathcal{P}_C(\check{\gamma})
	\triangleq & 2 (1-\epsilon) Q 
	\left( 
	\sqrt{g_{s,d} \gamma_{s,d} +  \varrho  \delta g_{r,d}|h_{s,r}|^2 \gamma_{r,d}}
	\right)   +  2(1-\epsilon) Q \left(
	\sqrt{ g_{s,d} \gamma_{s,d}} + \notag \right. \\ &  \left. \frac{\eta}{2}
	\frac{1}{\sqrt{  g_{s,d} \gamma_{s,d} }}
	\right) , \label{eq:PC}  \\
	\mathcal{P}_E(\check{\gamma}) 
	\triangleq &  
	\frac{2 \epsilon}{M-1}  Q \left(
	\sqrt{ g_{s,d} \gamma_{s,d}} - \frac{\eta}{2}
	\frac{1}{\sqrt{  g_{s,d} \gamma_{s,d} }}
	\right)  +  2 \epsilon 
	Q \left(
	\sqrt{ g_{s,d} \gamma_{s,d}}
	\right)  . \label{eq:PE}	
	\end{align}
\end{proposition}
Correspondingly, the overall average SER can be obtained by averaging over the channel gains as
\begin{align}
	\mathcal{P}_e = \int \mathcal{P}_{C}(\check{\gamma} )
	p_{\check{\gamma}} (\check{\gamma}) d \check{\gamma} + \int \mathcal{P}_{E}(\check{\gamma} )
	p_{\check{\gamma}} (\check{\gamma}) d \check{\gamma} ,
\end{align} where $p_{\check{\gamma}}(\check{\gamma})$ denotes the joint probability density function of $\check{\gamma}$. It is verified in Section \ref{sec:simu} that the proposed SER expression is quite accurate for not too low SNR values. Therefore, it is a good approximation for the actual SER of the system, and will be used for the SER analysis in the following sections.
%Further, adopting the observation that $ |y_{r,d}[k]-  y_{r,d}[k-1]x_s|^2  <   |y_{r,d}[k]-  y_{r,d}[k-1]x_s|^2 + \eta$ holds when $\eta > 0$, and exploiting the advantage that the $\min$ operations are exchangeable, the proposed detector is obtained as
%\begin{align}		
%\hat{x}_s  = \arg\min \left\{\min_{\mathbf{x} \in \mathcal{X}^2, x_s=x_r } m(\mathbf{x}), \min_{\mathbf{x} \in \mathcal{X}^2} m(\mathbf{x}) + \eta \right\}, \label{eq:NMLD-metric}
%\end{align}
%where
%$m(\mathbf{x}) \triangleq  \| \mathbf{y}-\mathbf{G}\mathbf{x}^T \|^2 
%$, $\mathbf{y}=[y_{s,d}[k], y_{r,d}[k]]^T$, $\mathbf{G} =  \diag \left(y_{s,d}[k-1],  y_{r,d}[k-1]   \right)
%$, and $\mathbf{x}=[x_s, x_r]$.
%%-------------------------------------------------------------------------
{ \color{black} 
	\subsection{Diversity Order Analysis}
\label{sec:div}
We first analyze the diversity order for the relay detection by examining the average SER $\epsilon $ in  \eqref{eq:SER-p2p-DM-2}, where $\bar{\gamma}_{s,r}^{ID} (\varrho) = \frac{2(1-\varrho) T_s L_{s,r} }{2-\varrho} \bar{\gamma}_{s,r}$ by substituting $N_{sr,1}=N_{sr,2}=N_0/2$. After some manipulations (provided in Appendix \ref{app:proof-SER-ave}), we have  
\begin{align}  \label{eq:SER-p2p-div}
\lim\limits_{\bar{\gamma}_{s,r} \to \infty} \epsilon =  
\begin{cases} 
\frac{1}{4 T_s L_{s,r}} \frac{ 2-\varrho }{(1-\varrho) \bar{\gamma}_{s,r} }, & M = 2,\\
\frac{1.03}{4} \sqrt{\frac{1+\cos \frac{\pi}{M}}{2 \cos \frac{\pi}{M}}} \frac{1}{(1-\cos \frac{\pi}{M})  T_s L_{s,r} } \frac{ 2-\varrho }{(1-\varrho) \bar{\gamma}_{s,r} }, & M > 2,
\end{cases}
\end{align}
which decays with  $\frac{1}{\bar{\gamma}_{s,r}}$ for all $0 < \varrho <1$, achieving the full diversity order of one. We can write $\lim\limits_{\bar{\gamma}_{s,r} \to \infty} \epsilon = c_0 \frac{ 2-\varrho }{(1-\varrho) \bar{\gamma}_{s,r} }$ with some constant $c_0$, and then we have $\lim\limits_{\bar{\gamma}_{s,r} \to \infty}  \eta = \ln \frac{1}{\epsilon}$. Note that such analysis was not conducted in \cite{dm-df-ser-1-liu2015energy}.

We now analyze the SER expressions in \textbf{Proposition \ref{prop: 1}} for the entire network. To obtain a closed-form average  SER, we derive in Appendix \ref{app:proof-SER-ave} that $\mathcal{P}_e \approx \mathcal{P}_C + \mathcal{P}_E$ and 
\begin{align}
\mathcal{P}_C = & (1-\epsilon)(Z_1 + Z_2), \label{eq:PC-ave}\\
\mathcal{P}_E = & \frac{\epsilon  Z_3}{M-1}  +  \frac{\epsilon }{ g_{s,d} \bar{\gamma}_{s,d}+2},  \label{eq:PE-ave}
\end{align}
where $Z_1 =  a_1 \sqrt{2\eta} \exp (-2b_1 \eta) $, $Z_2 =    \frac{a_2}{\varrho} \ln \left(
1+b_2 \varrho \right)$, $Z_3 =  \exp (\eta) Z_1$, and some auxiliary variables not related to $\varrho$ are defined as $a_1 = \frac{\sqrt{\pi}(2 g_{s,d})^{-\frac{1}{4}}}{4 \bar{\gamma}_{s,d}} \left( g_{s,d}/2+\bar{\gamma}_{s,d}^{-1} \right)^{-\frac{3}{4}}$, $a_2 = \frac{ 2 \frac{P_s}{N_0}}{\delta g_{r,d} ( g_{s,d} \bar{\gamma}_{s,d}+2) \bar{\gamma}_{s,r} \bar{\gamma}_{r,d}}$, $b_1 = \frac{1}{4} + \frac{ \sqrt{ g_{s,d}/2+\bar{\gamma}_{s,d}^{-1}}}{2  \sqrt{2 g_{s,d}}}$, and $b_2 = \frac{ \delta g_{r,d}  \bar{\gamma}_{s,r} \bar{\gamma}_{r,d}}{2 \frac{P_s}{N_0}}$. 

Since 
\begin{align}
	\lim\limits_{\bar{\gamma}_{s,r} \to \infty, \bar{\gamma}_{s,d} \to \infty} \ (1-\epsilon) Z_1 = &  c_1 \frac{\sqrt{\ln \bar{\gamma}_{s,r}} }{\bar{\gamma}_{s,r} \bar{\gamma}_{s,d}}, \\
	\lim\limits_{\bar{\gamma}_{s,r} \to \infty, \bar{\gamma}_{s,d} \to \infty, \bar{\gamma}_{r,d} \to \infty} \ (1-\epsilon) Z_2 = &  \frac{ c_{2,1} }{\bar{\gamma}_{s,r} \bar{\gamma}_{s,d} \bar{\gamma}_{r,d} \varrho}  \ln \frac{ \bar{\gamma}_{s,r} \bar{\gamma}_{r,d} \varrho }{c_{2,2}}, \\
	\lim\limits_{\bar{\gamma}_{s,r} \to \infty, \bar{\gamma}_{s,d} \to \infty} \ \epsilon Z_3 = &  c_3 \frac{ 2-\varrho }{(1-\varrho) \bar{\gamma}_{s,r} } \frac{\sqrt{\ln \bar{\gamma}_{s,r}}}{ \bar{\gamma}_{s,d} } , \\
	\lim\limits_{\bar{\gamma}_{s,r} \to \infty, \bar{\gamma}_{s,d} \to \infty} \ \frac{\epsilon }{ g_{s,d} \bar{\gamma}_{s,d}+2} = &  c_4 \frac{ 2-\varrho }{(1-\varrho) \bar{\gamma}_{s,r} \bar{\gamma}_{s,d} }, 	 
\end{align}
with some constants $\{ c_1, c_{2,1}, c_{2,2}, c_3, c_4 \}$, we have 
\begin{align}
	\lim\limits_{\bar{\gamma}_{s,r} = \bar{\gamma}_{s,d} = \bar{\gamma}_{r,d} = \bar{\gamma} \to \infty} \ \left( \mathcal{P}_C + \mathcal{P}_E \right) \propto \frac{1}{\bar{\gamma}^2},
\end{align}
where $\bar{\gamma}_{s,r} = \bar{\gamma}_{s,d} = \bar{\gamma}_{r,d} = \bar{\gamma}$ is used without loss of generality and $\varrho$ is assumed to be finite. Based on the above, we can see that $\mathcal{P}_C + \mathcal{P}_E$ decays at the rate $\frac{1}{\bar{\gamma}^2}$ for sufficiently high SNR, and therefore the proposed detector achieves the full diversity order, which is two.
}
%%--------------------------------------------------------------------------------------------------------
\section{Optimized Power Splitting Ratio}
\label{sec:ana}
In this section, we analyze the SER performance in terms of $\varrho$ and develop two methods to estimate the optimal value of $\varrho$.
%%--------------------------------
\subsection{SER Performance Trade-off}
Useful insights can be drawn from the proposed approximate SER expression in various aspects. As an example, we present the trade-off between the conditional SERs of the two scenarios where the relay detects correctly and wrongly as a function of $\varrho$ in this subsection.

The EH relay system should take advantage of a $S-R$ link with good quality, and therefore here we assume sufficiently high average SNR of the $S-R$ link, then there is $\epsilon  \rightarrow 0 $ and $\eta \approx  \ln \frac{1}{\epsilon} \rightarrow \infty $. In this case, $\mathcal{P}_C (\check{\gamma})$ and $\mathcal{P}_E (\check{\gamma})$ can be further approximated using the dominating terms $\tilde{\mathcal{P}}_C(\check{\gamma})$ and $\tilde{\mathcal{P}}_E(\check{\gamma})$, respectively, as 
\begin{align}
\tilde{\mathcal{P}}_C (\check{\gamma}) 
\triangleq & 	2(1-\epsilon) Q 
\left( 
\sqrt{g_{s,d} \gamma_{s,d} +  \varrho  \delta g_{r,d}|h_{s,r}|^2 \gamma_{r,d}}
\right) , \\
\tilde{\mathcal{P}}_E  (  \check{\gamma} ) 
\triangleq &    
\frac{2\epsilon}{M-1}  Q \left(
\sqrt{ g_{s,d} \gamma_{s,d}} - \frac{\eta}{2}
\frac{1}{\sqrt{  g_{s,d} \gamma_{s,d} }}
\right)  .
\end{align}

We want to emphasize that both $\epsilon$ and $\eta$ are functions of $\varrho$ {\color{black}(c.f. \eqref{eq:SER-p2p-DM-2})}, and therefore $\tilde{\mathcal{P}}_C (\check{\gamma})$ and $\tilde{\mathcal{P}}_E (\check{\gamma})$ are functions of $\varrho$. By proving in Appendix \ref{appen-mono} that $\tilde{\mathcal{P}}_C( \check{\gamma} ) $ and $ \tilde{\mathcal{P}}_E(  \check{\gamma} ) $ are monotonically decreasing and increasing in $\varrho$, respectively, we can see that there exists a trade-off between the conditional SERs of the aforementioned two scenarios. One possible explanation is that increasing the PS ratio $\varrho$ will increase the relay transmission power $P_r$. For scenario one where the detection at $R$ is correct, the reliability of the overall $S-R-D$ link is increased, and the error probability decreases, while for scenario two where the detection at $R$ is wrong, increasing $P_r$ encourages error propagation, which in turn undermines the network reliability and increases the error probability. The results suggest that a good trade-off can potentially be achieved by adjusting $\varrho$. 
%\begin{align}
%& \frac{\frac{1}{ 2 \sqrt{2 \pi g_{s,d} \gamma_{s,d}} } \exp \left(-
%	\frac{z_0^2}{2}
%	\right) }{1-\frac{1}{12} \exp \left(-\frac{z_0^2}{2}\right) - \frac{1}{4} \exp \left(-\frac{2}{3}z_0^2 \right)} \notag \\ \approx & \frac{1}{ 2 \sqrt{2 \pi g_{s,d} \gamma_{s,d}} } \exp \left(-
%\frac{z_0^2}{2}
%\right) \rightarrow 0
%\end{align}
\begin{comment}
On the other hand, this tradeoff still exists averaging over the Rayleigh fading channel gains. The justifications are:

For $\tilde{\mathcal{P}}_C( \check{\gamma} )$, it can be seen that averaging over the channel gains will not change its monotonicity in $\varrho$ since $\varrho$ only serves as a positive scaling factor of the gain term $|h_{s,r}|^2 \gamma_{r,d}$. For $\tilde{\mathcal{P}}_E( \check{\gamma} )$, since it is positive, monotonically increases in $\varrho$, and $\varrho$ does not affect the channel gain $\gamma_{s,d}$ in any way, averaging over $\gamma_{s,d}$ will not change the monotonically, either. 

Therefore, we conclude that $\int \tilde{\mathcal{P}}_C( \check{\gamma} ) p_{\check{\gamma}} (\check{\gamma}) d ( \check{\gamma} )$ and $\int \tilde{\mathcal{P}}_E( \check{\gamma} ) p_{\check{\gamma}} (\check{\gamma}) d ( \check{\gamma} )$ also monotonically decrease and increase in $\varrho$, respectively with $p_{\check{\gamma}}(\check{\gamma})$ as the joint probability density function of $\check{\gamma}$. Tradeoff still exists following the earlier illustrations.
\end{comment}
%%--------------------------------------------------------
\subsection{Optimized Power Splitting Ratio}
Simulation results in \cite{dm-df-ser-1-liu2015energy} show that there is an optimal value of $\varrho$ that minimizes the SER ({\color{black}see also our simulation results in Section \ref{sec:simu}}). However, the task of finding the closed-form expression for this optimal value appears intractable. To address this issue, in this subsection, we propose two methods to estimate this optimal PS ratio off-line numerically. %  without the need for performing Monte-Carlo simulations. % 

%Our simulation results in Section \ref{sec:simu} also show that there is a unique value of the PS ratio $\varrho$ that minimizes the SER, same as the results in \cite{dm-df-ser-1-liu2015energy}. 
%One prominent design limitation for such non-coherent system is that no instantaneous CSI is available at any node, therefore only the average CSI can be used.

%
%We assume symmetric links with $\bar{\gamma}_{I,J} = \bar{\gamma}, I,J \in \{s,r,d\}$ with the average SNR $\bar{\gamma} \triangleq \frac{P_s}{N_0}$, and asymmetric case can be studied through adjusting the path loss $L_{I,J}$.

One method is by calculating the minimum of the average approximate SER expression, i.e., $\mathcal{P}_e$, using software packages such as cvx with MATLAB; see \cite{cvx-grant2014cvx} for details. This method is straightforward. However, double integral calculations are needed. To save the computational effort, we propose a second method by equivalently examining the zero of its derivative with respect to $\varrho$, i.e., the zero of $\frac{\partial \mathcal{P}_C}{\partial \varrho} + \frac{\partial \mathcal{P}_E}{\partial \varrho}$. 
Based on the chain rule of derivative, we have
\begin{small}
	\begin{align}
	\frac{\partial \mathcal{P}_C}{\partial \varrho} = & - \frac{\partial \epsilon}{\partial \varrho} (Z_1 + Z_2) + (1-\epsilon) 
	\left(
	\frac{\partial Z_1 }{\partial \eta } \frac{\partial \eta }{\partial \epsilon} \frac{\partial \epsilon}{\partial \varrho} + 
	\frac{\partial Z_2 }{\partial \varrho }
	\right), \notag  \\
	\frac{\partial \mathcal{P}_E}{\partial \varrho} = & \frac{Z_3}{M-1} \frac{\partial \epsilon}{\partial \varrho}  + \frac{\epsilon}{M-1}   \frac{\partial Z_3 }{\partial \eta } \frac{\partial \eta }{\partial \epsilon} \frac{\partial \epsilon}{\partial \varrho} + \frac{1 }{ g_{s,d} \bar{\gamma}_{s,d}+2} \frac{\partial \epsilon}{\partial \varrho} \notag \\
	= & \left( \frac{\epsilon }{M-1}  \frac{\partial Z_3 }{\partial \eta } \frac{\partial \eta }{\partial \epsilon}  + \frac{ Z_3}{M-1}  + \frac{1 }{ g_{s,d} \bar{\gamma}_{s,d}+2} \right)  \frac{\partial \epsilon}{\partial \rho(  \varrho )}
	\frac{\partial \rho(  \varrho ) }{\partial \varrho}, \notag
	\end{align}
\end{small}where the function $\rho(  \varrho )$ is defined as $\rho(  \varrho ) \triangleq \frac{1-\varrho}{ 2-\varrho }$. By further substitutions, we have that the derivative, i.e., $\frac{\partial \mathcal{P}_C}{\partial \varrho} + \frac{\partial \mathcal{P}_E}{\partial \varrho}$, can be obtained in closed-form as
% \begin{small}
\begin{align}
\frac{\partial \mathcal{P}_C}{\partial \varrho} = &  \left[ a_1 \sqrt{2\eta} \exp (-2b_1 \eta) + \frac{a_2}{\varrho} \ln \left(
1+b_2 \varrho \right)
\right]  
\sqrt{  \frac{ 1 + \rho(  \varrho )b_3 }{\rho(  \varrho ) b_3}  } \frac{ -a_3b_3 }{ 2 (1+ \rho(  \varrho ) b_3)^2(2-\varrho)^2}  + 	\frac{ a_1}{ \epsilon }  \notag \\
&  \exp (-2b_1 \eta)
\left[ -   b_1 \sqrt{2\eta}  +  \frac{1}{2\sqrt{2\eta}}   	\right]
\sqrt{  \frac{ 1 + \rho(  \varrho )b_3 }{\rho(  \varrho ) b_3}  }  \frac{ -a_3b_3 }{ (1+ \rho(  \varrho ) b_3)^2(2-\varrho)^2}  + 
(1-\epsilon) \notag \\
& 
\left[
-\frac{a_2}{\varrho^2} \ln \left(
1+b_2 \varrho \right) + \frac{a_2 b_2}{\varrho(1+b_2 \varrho)} 
\right]  \label{eq:deriv-1},	\\
\frac{\partial \mathcal{P}_E}{\partial \varrho} = &  \left[ \frac{1 }{ g_{s,d} \bar{\gamma}_{s,d}+2} + \frac{a_1}{M-1}   \exp (\eta) \exp (-2b_1 \eta)  \left[
\left( \frac{ \sqrt{2\eta} }{ 2 } - b_1 \sqrt{2\eta} + \frac{1}{ 2 \sqrt{2\eta} } 
\right)  \frac{-2}{  (1-\epsilon)} + \right.\right. \notag \\
& \left.\left. \sqrt{2\eta}
\right]
\right]  \sqrt{  \frac{ 1 + \rho(  \varrho )b_3 }{\rho(  \varrho ) b_3}  } \frac{ a_3b_3 }{ 2 (1+ \rho(  \varrho ) b_3)^2 (2-\varrho)^2}  \label{eq:deriv-2},
\end{align}
% \end{small}
where $a_3 = 1.03 \sqrt{\frac{1+\cos \frac{\pi}{M}}{2 \cos \frac{\pi}{M}}} $ and $b_3 = 2 (1-\cos \frac{\pi}{M}) T_s L_{s,r} \bar{\gamma}_{s,r}$ are not related to $\varrho$.

Numerical results in Section \ref{sec:simu} show that the estimated derivative is monotonically increasing with $\varrho$ and has a unique zero, which well matches the minimum of the simulated SER.

%%--------------------------------------------------------------------------------------------------------
{ \color{black}
	\section{Extended Results for the Time Switching Protocol} \label{sec:TS}
In this section, we extend the proposed detector and the SER analysis results for another widely adopted EH protocol, i.e., the TS protocol. 	For simplicity, we will reuse some notations in Section \ref{sec:sys-mod}.
%%------------------------------------------------------
\subsubsection{Proposed Detector at the Destination}  \label{sec:sys-TS}
As shown in Fig.~\ref{fig:TS}, for the TS protocol, the WPT and WIT processes are performed in a time-division fashion, with $\alpha T$ duration for WPT and $(1-\alpha) T$ for WIT, where $0 < \alpha < 1$ is called the TS ratio. 
In the WPT process, the harvested energy at $R$ is $E_r = \delta P_s L_{s,r} |h_{s,r}|^2 \alpha T$. 
\begin{figure}[t]
	\centering
	\includegraphics[width=1\textwidth]{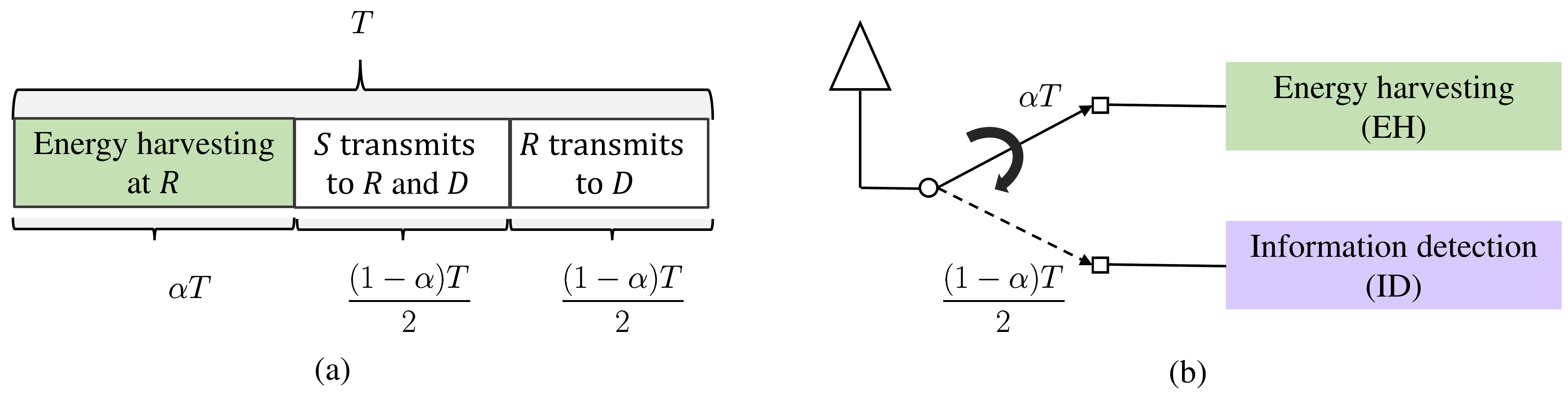}%{PS-channel-model}
	\caption{\color{black}Illustration for  the TS protocol. (a) The WPT (green box) and WIT (white boxes) processes.  (b) Block diagram of the relay receiver architecture.}
	\label{fig:TS}
\end{figure}

In the WIT process, one symbol transmission consists of two time slots each with duration $T_s = \frac{(1-\alpha)T}{2}$.  In a unified form, the received signals at $R$ for ID and at $D$ are
\begin{align}
y_{I,J}[k] = & \sqrt{P_I T_s L_{I,J}} h_{I,J} u_I[k] + v_{IJ,1}[k] + v_{IJ,2}[k], \label{eq:y-TS}
\end{align}
where $I \in \{s,r\}$ denotes the transmitting node, $J \in \{r,d\}$ denotes the corresponding receiving node, and 
$P_r$ denotes the transmission power at $R$ obtained from WPT, given by
\begin{align}
P_r = P_r(\alpha, h_{s,r}) = \frac{E_r}{\frac{(1-\alpha)T}{2}} = \frac{2 \delta P_s L_{s,r} |h_{s,r}|^2 \alpha}{1-\alpha}. \label{eq:Pr-TS}
\end{align}

Following the similar derivations as in Section \ref{sec:det}, the proposed detector for using the TS protocol  is performed based on \eqref{eq:det}, where $ y_{r,d}$ is given in \eqref{eq:y-TS} and is a function of $\alpha$ and $h_{s,r}$, $\eta \triangleq \ln\frac{(1-\epsilon )(M-1)}{\epsilon } $, and $\epsilon$ is given by (c.f.  \cite[eq. (33)]{dm-df-ser-1-liu2015energy})
\begin{align} \label{eq:SER-p2p-DM-TS}
\epsilon  \triangleq  
\begin{cases} 
\frac{ 1 }{2 \left[1+ \bar{\gamma}_{s,r}^{ID}  \right]}, & M = 2,\\
1.03 \sqrt{\frac{1+\cos \frac{\pi}{M}}{2 \cos \frac{\pi}{M}}}   \left[ 1-
\sqrt{\frac{ (1-\cos \frac{\pi}{M})  \bar{\gamma}_{s,r}^{ID}  }{1+  (1-\cos \frac{\pi}{M}) \bar{\gamma}_{s,r}^{ID}  }} \
\right], & M > 2,
\end{cases}
\end{align} where $\bar{\gamma}_{s,r}^{ID}  = \frac{ T_s  P_s L_{s,r} }{ N_{sr,1} + N_{sr,2}}$. Clearly, the detection complexity is still $\mathcal{O}(M)$.

%%------------------------------------------------------
\subsubsection{Approximate SER for the Detector} 
The approximate SER expression is presented in \textbf{Proposition \ref{prop: 2}} and the proof is given in Appendix \ref{proof-prop-2}.
By calculating the minimum of the average approximate SER, i.e., $\int \mathcal{P}_{C}(\check{\gamma} )
p_{\check{\gamma}} (\check{\gamma}) d \check{\gamma} + \int \mathcal{P}_{E}(\check{\gamma} )
p_{\check{\gamma}} (\check{\gamma}) d \check{\gamma}$, we can estimate the optimal TS ratio, as will be shown in Section \ref{sec:simu-TS}.
\begin{proposition}\label{prop: 2}
	For the SWIPT-enabled DDF single-relay network adopting the TS protocol, the overall SER of the proposed near-optimal detector is approximately equal to $  \mathcal{P}_C (\check{\gamma}) + \mathcal{P}_E  (  \check{\gamma} )$ for $M>2$, and $ \frac{1}{2} \mathcal{P}_C (\check{\gamma}) + \frac{1}{2} \mathcal{P}_E  (  \check{\gamma} )$ for $M=2$, where
	\begin{align}
	\mathcal{P}_C(\check{\gamma})
	\triangleq & 
	2(1-\epsilon) Q 
	\left(
	\sqrt{  g_{s,d} \gamma_{s,d} +   \frac{2\alpha}{1-\alpha}\delta g_{r,d} |h_{s,r}|^2 \gamma_{r,d}}
	\right) + 2(1-\epsilon) 
	Q 	\left( 
	\sqrt{ g_{s,d} \gamma_{s,d}} + \right. \notag \\ & \left. \frac{\eta}{2} \frac{ 1 }{   \sqrt{ g_{s,d} \gamma_{s,d}} } 
	\right) 	 , \label{eq:PC-TS}  \\
	\mathcal{P}_E(\check{\gamma}) 
	\triangleq &  \frac{2 \epsilon}{M-1}  Q \left(
	\sqrt{  g_{s,d} \gamma_{s,d} } - \frac{\eta}{2}
	\frac{1}{\sqrt{ g_{s,d} \gamma_{s,d} }}
	\right)  + 
	2 \epsilon 
	Q \left(
	\sqrt{  g_{s,d} \gamma_{s,d} } 
	\right)  
	. \label{eq:PE-TS}	
	\end{align}
\end{proposition}

To obtain a closed-form average  SER, similarly to the derivation for the PS case (details are omitted here), we can obtain $\mathcal{P}_e \approx \mathcal{P}_C + \mathcal{P}_E$ and 
\begin{align}
\mathcal{P}_C = & (1-\epsilon)(Z_1 + Z_2), \label{eq:PC-ave-TS}\\
\mathcal{P}_E = & \frac{\epsilon  Z_3}{M-1}  +  \frac{\epsilon }{ g_{s,d} \bar{\gamma}_{s,d}+2},  \label{eq:PE-ave-TS}
\end{align}
where $Z_2 =    \frac{a_2(1-\alpha)}{2\alpha} \ln \left(
1+b_2 \frac{2\alpha}{1-\alpha} \right)$ and the expressions for the other variables are the same as those in the PS case. After some manipulations, it can be proved that 
\begin{align}
\lim\limits_{\bar{\gamma}_{s,r} = \bar{\gamma}_{s,d} = \bar{\gamma}_{r,d} = \bar{\gamma} \to \infty} \ \left( \mathcal{P}_C + \mathcal{P}_E \right) \propto \frac{1}{\bar{\gamma}^2},
\end{align}
where $\alpha$ is assumed to be finite.
Therefore, the proposed detector achieves the full diversity order of two (the same conclusion as in Section \ref{sec:div}).

%%------------------------------

}

%%--------------------------------------------------------------------------------------------------------
\section{Numerical and Simulation Results}
\label{sec:simu}
In this section, numerical and simulation results are presented for evaluation. The PS protocol is adopted at the relay without otherwise stated. The near-optimal SER performance of the proposed detector and the accuracy of the SER expression are shown. The accuracies of the proposed two methods for estimating the optimal PS ratio are also validated.  The EH efficiency is set to $\delta=0.6$ without otherwise stated, and we use the bounded path-loss model as $L_{I,J}=\frac{1}{1+d_{I,J}^{2.7}}$ to ensure that the path-loss is strictly larger than one. The distances between the nodes are set as $d_{s,d}=3$, and $d_{s,r}=d_{r,d}=1.5$ by default. The SER performance is parameterized by the transmitter SNR defined as SNR $\triangleq P_s/N_0$. The transmission rate is set as $R=\ln M$ for simplicity, which only depends on the modulation size $M$.

%%-------------------------%%
\subsection{ \color{black} SER Performance Comparison for Different Modulation Size $M$ }
\begin{figure}[t]
	\centering
	\includegraphics[width=0.6\textwidth]{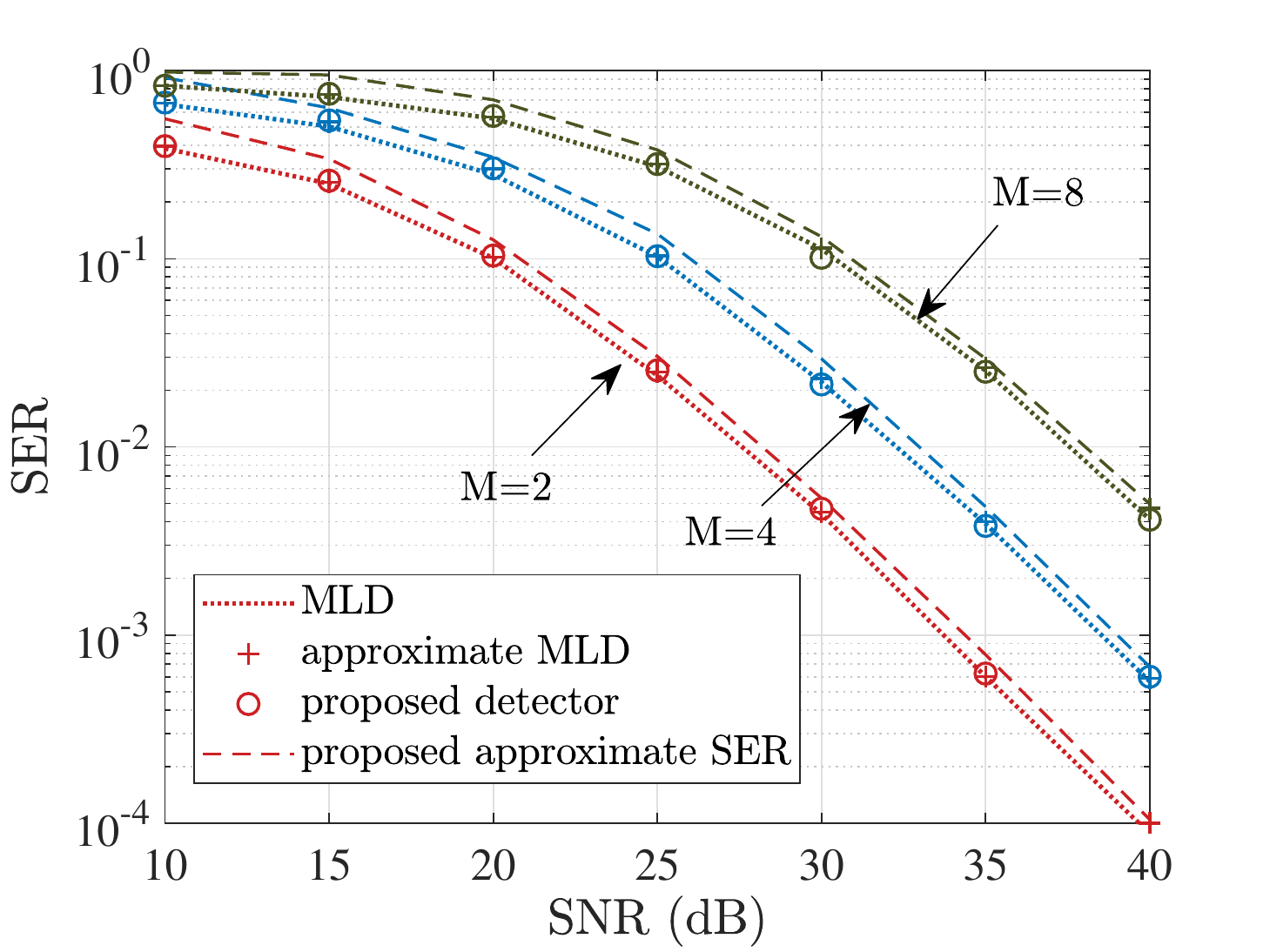}
	\caption{\color{black}SER performance comparison of different detectors and the proposed approximate SER expression with respect to the SNR (dB) when $\varrho=0.8$ and $\delta=0.6$, for the SWIPT-enabled DDF relay networks with $M$-DPSK.}
	\label{fig:SER-2}
\end{figure}

\begin{figure}[t] % \usepackage{subfigure}
	\centering
	\subfigure[$M=2$, SNR = $30$ dB ]{
		\label{fig:SER-PS-M2}
		\includegraphics[width=0.6\textwidth]{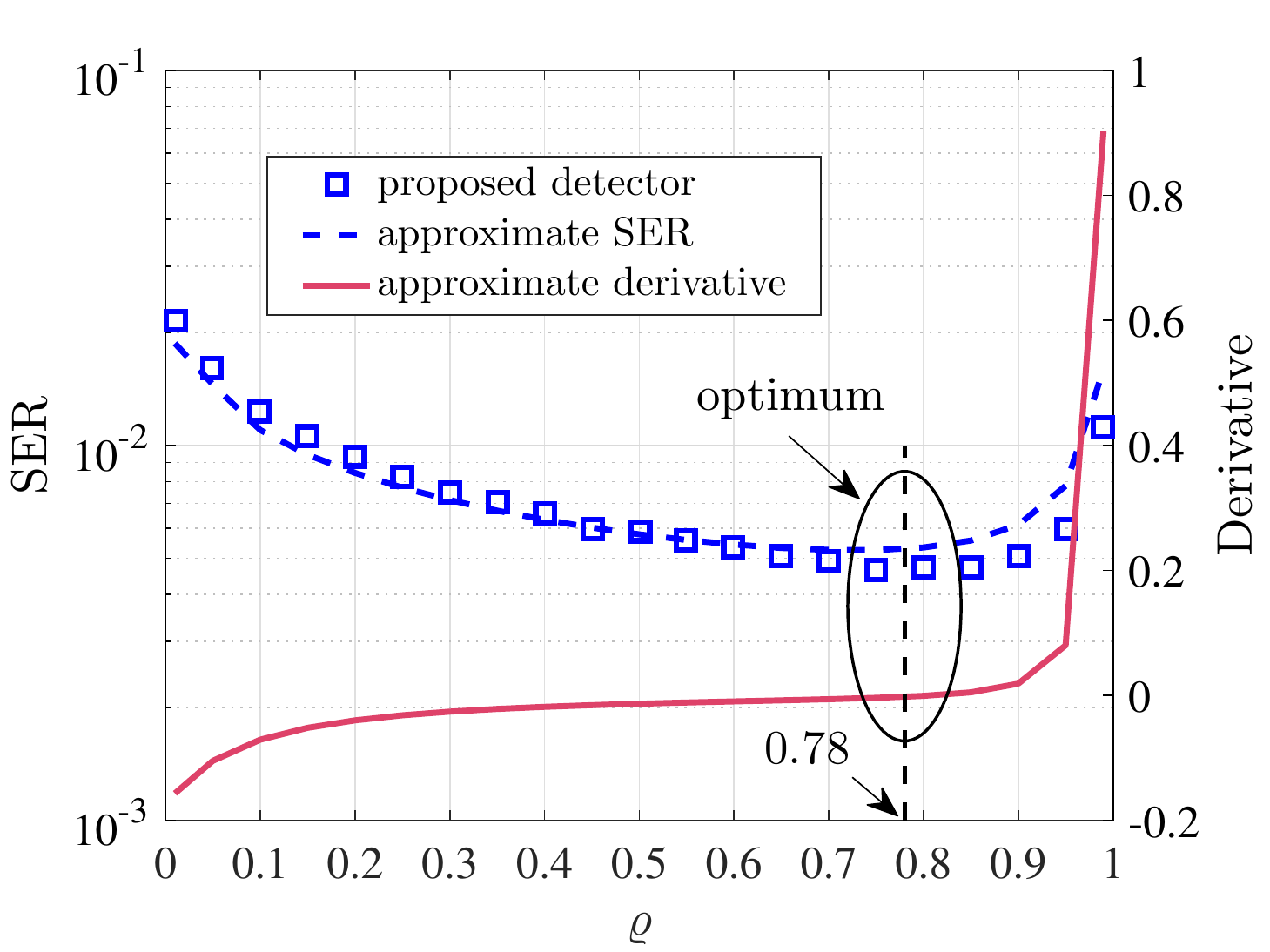}}
	%	\hspace{0.01\linewidth}
	\subfigure[$M=8$, SNR = $40$ dB]{
		\label{fig:SER-PS-M8}
		\includegraphics[width=0.6\textwidth]{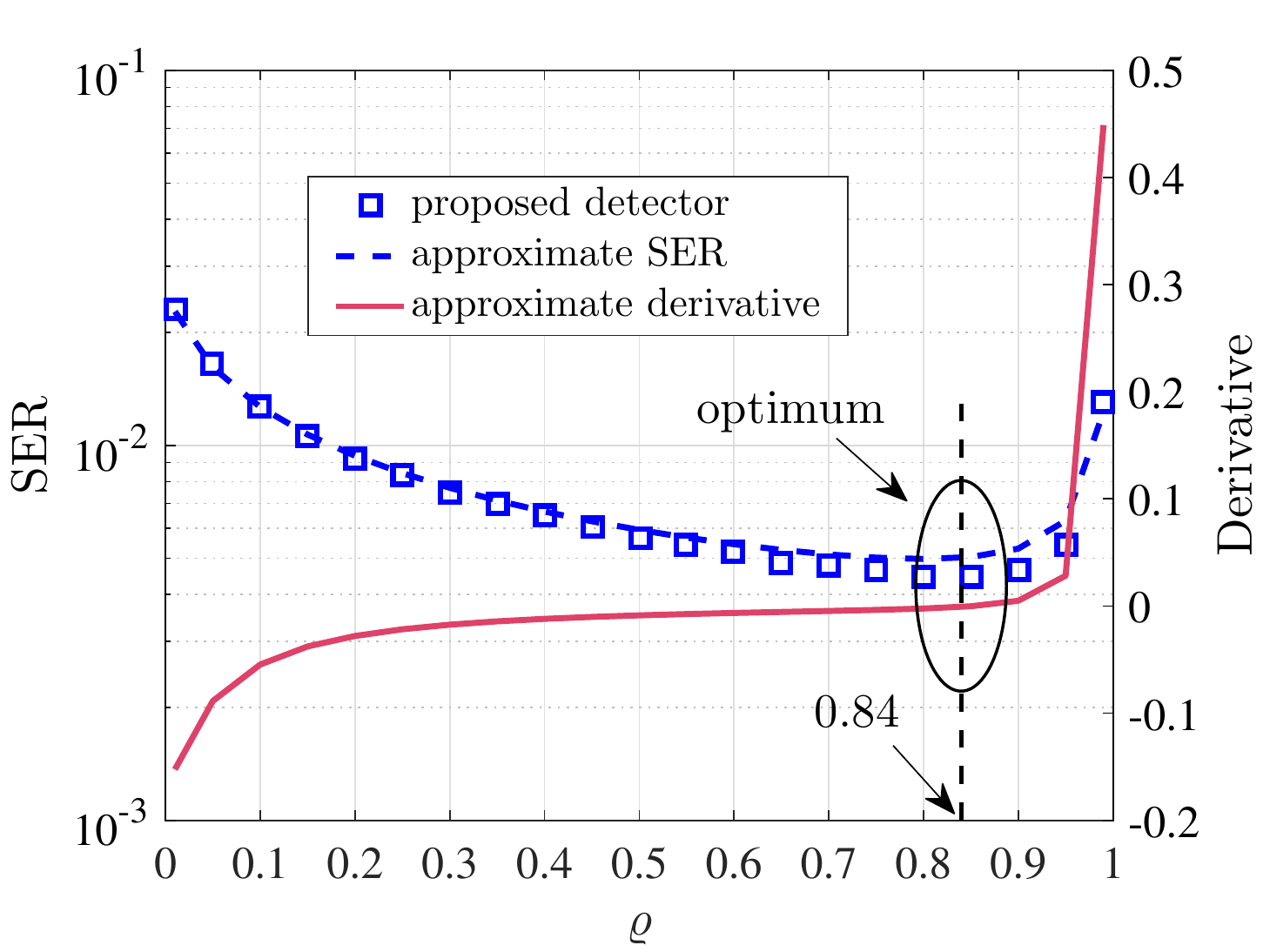}}
	\caption{The simulated SER of the proposed detector and the proposed approximate SER and derivative with respect to the PS ratio $\varrho$, when $\delta=0.6$, for the SWIPT-enabled DDF relay networks with $M$-DPSK.}
	\label{fig:SER-PS-opt}
\end{figure}

%%-------------------------%%
Fig.~\ref{fig:SER-2} compares the SER performances of our detector and the state-of-the-art {\color{black} MLD and} approximate MLD \cite{dm-df-ser-1-liu2015energy} for various transmission rates (modulation sizes) with respect to the SNR (dB), when $\varrho=0.8$. The proposed approximate SER expression is also simulated to show its accuracy. It can be seen that the {\color{black}three} detectors show similar SER performance, and {\color{black}all} achieve the full diversity order of two. Because it has been verified in \cite{dm-df-ser-1-liu2015energy} that {\color{black}the MLD can  characterize the optimal performance for ID in this SWIPT-enabled DDF relay network,}  the results verify that our detector is near-optimal. It is also notable that the approximate SER is quite accurate for not too low SNR values.

\subsection{ \color{black} Validation of the Optimized Power Splitting Ratio}
To validate the accuracy of the two methods proposed  in Section \ref{sec:ana}, which estimate the optimal PS ratio $\varrho$,  Fig.~\ref{fig:SER-PS-opt} shows the simulated SER of our detector, and approximate SER and derivative, for DBPSK with $R=1$ (bps) at SNR = $30$ dB and $8$-DPSK with $R=3$ (bps) at SNR = $40$ dB, with respect to $\varrho$. To show the SER and the derivative simultaneously, double y-axes is used with the SER value on the left y-axis and the derivative value on the right. There are several observations that can be made from Fig.~\ref{fig:SER-PS-opt}. The first is that the simulated SER has a unique minimum. The second is that the proposed approximate SER expression is quite accurate for all $0 < \varrho < 1$ considered, and also shows a unique minimum. The third is that the derivative is monotonically increasing from negative to positive with $\varrho$, which suggests that the proposed approximate SER expression is convex in $\varrho$. Most notably, it can be seen that the minimums of the simulated and approximate SERs and the zero of the approximate derivative are consistent to the second decimal digit. The optimal values are  $0.78$ and $0.84$, respectively, for $M=2$ and $8$. Therefore, both the approximate SER and derivative can be used to estimate the optimal PS ratio accurately. 

\begin{figure}[h]
	\centering
	\includegraphics[width=0.6\textwidth]{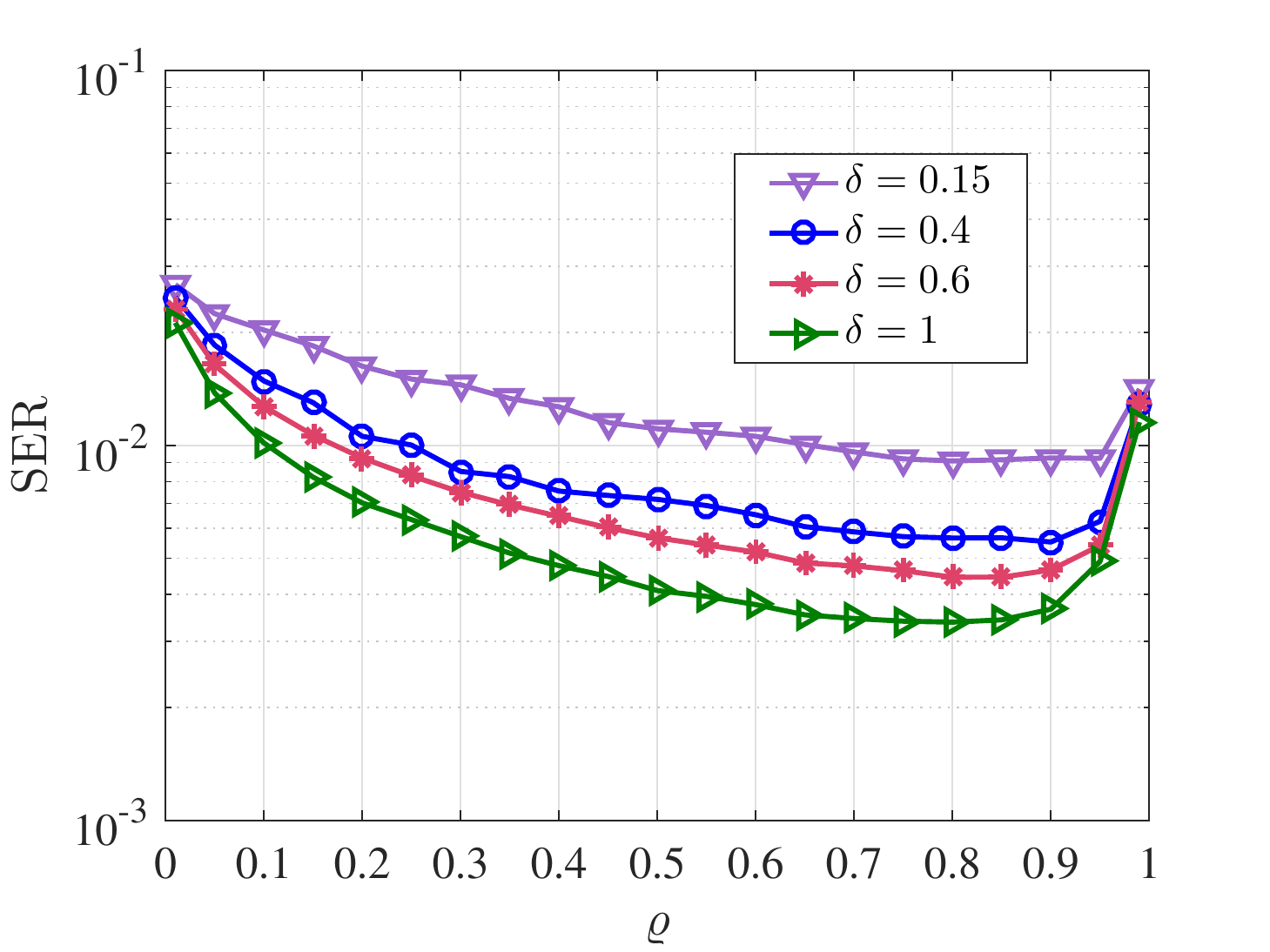}
	\caption{SER performance comparison for different EH efficiency $\delta$ with respect to the PS ratio $\varrho$, using the proposed detector with $8$-DPSK at SNR $= 40$ dB.}
	\label{fig:SER-effi}
\end{figure}

%%-------------------------%%
\subsection{ \color{black} Impacts of the Energy Harvesting Efficiency $\delta$ and Distance $d_{r,d}$}

Fig.~\ref{fig:SER-effi} compares the simulated SER for different values of the EH efficiency $\delta=0.15, 0.4, 0.6$, and $1$ using our detector. The modulation is $8$-DPSK and SNR $= 40$ dB. It can be observed that for a fixed $0 < \varrho < 1$, the SER decreases with $\delta$. This is possibly because as $\delta$ increases, $R$ is capable of harvesting more power from the same received signals, and therefore the overall SER performance is improved. It is also notable that increasing the EH efficiency $\delta$ will shift the optimal value of $\varrho$ to left. An interpretation of this is that as $\delta$ increases, $R$ becomes more energy efficient, and therefore a smaller value of $\varrho$ is needed to maintain the same reliability as that of the previous. 

{ \color{black}

Fig.~\ref{fig:SER-drd} compares the simulated SER for different values of the distance $d_{r,d}=1, 1.5$, and $2$ using our detector for  $d_{s,d}=3$,  $d_{s,r}=d_{s,d}- d_{r,d}$, and $\delta=1$. The modulation is $8$-DPSK and SNR $= 40$ dB.
It can be seen that the optimal value of $\varrho$ increases as $d_{r,d}$  increases.  This is possibly because as $d_{r,d}$ increases,  the $S-R$ link becomes better while the $R-D$ link becomes worse. Then, less power is needed at $R$ for ID to achieve the same reliability as that of the previous.
\begin{figure}[t]
	\centering
	\includegraphics[width=0.6\textwidth]{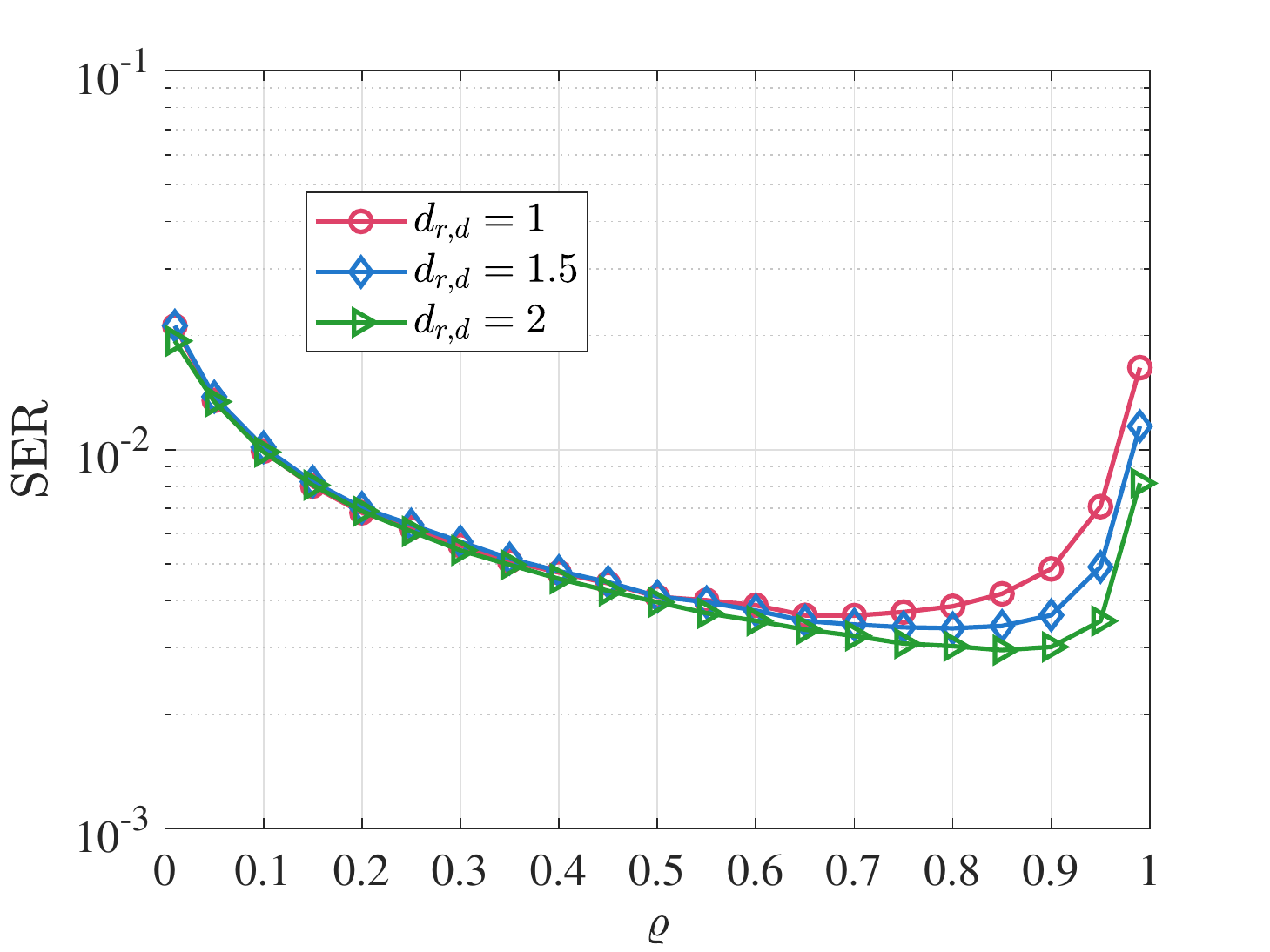}
	\caption{ \color{black} SER performance comparison for different $R-D$ link distance  $d_{r,d}$ with respect to the PS ratio $\varrho$, where  $d_{s,d}=3$,  $d_{s,r}=d_{s,d}- d_{r,d}$, and $\delta=1$, using the proposed detector with $8$-DPSK at SNR $= 40$ dB.}
	\label{fig:SER-drd}
\end{figure}
}
%%-------------------------%%
%\newpage
\subsection{ \color{black} SER Performance for Adopting the TS Protocol }
\label{sec:simu-TS}
\begin{figure}[t]
	\centering
	\includegraphics[width=0.6\textwidth]{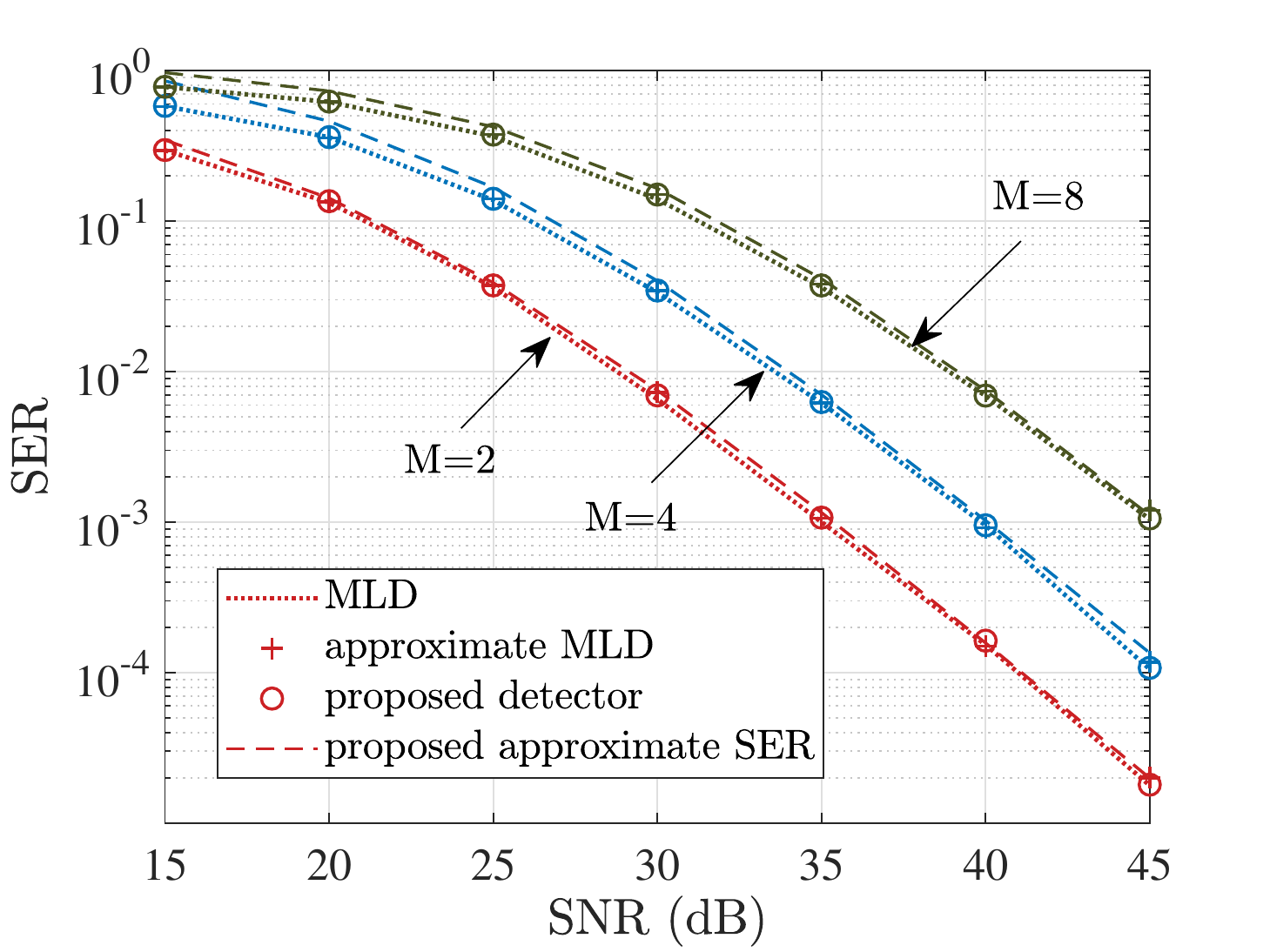}
	\caption{\color{black}SER performance comparison of different detectors and the proposed approximate SER  expression with respect to the SNR (dB) when $\alpha=0.4$ and $\delta=0.6$ adopting the TS protocol, for the SWIPT-enabled DDF relay networks with $M$-DPSK.}
	\label{fig:SER-TS}
\end{figure}

{ \color{black}
In this subsection, we verify the results for the TS protocol in Section \ref{sec:TS}. 

Fig.~\ref{fig:SER-TS} compares the SER performances of the proposed detector and the state-of-the-art detectors for various modulation sizes, when $\alpha=0.4$ adopting the TS protocol at the relay. The proposed SER expression is also shown. It can be seen that our detector performs very} { \color{black}close to the optimal MLD and achieves the full diversity order of two. The SER expression is shown to be accurate for various $M$.}

\begin{figure}[t]
	\centering
	\includegraphics[width=0.6\textwidth]{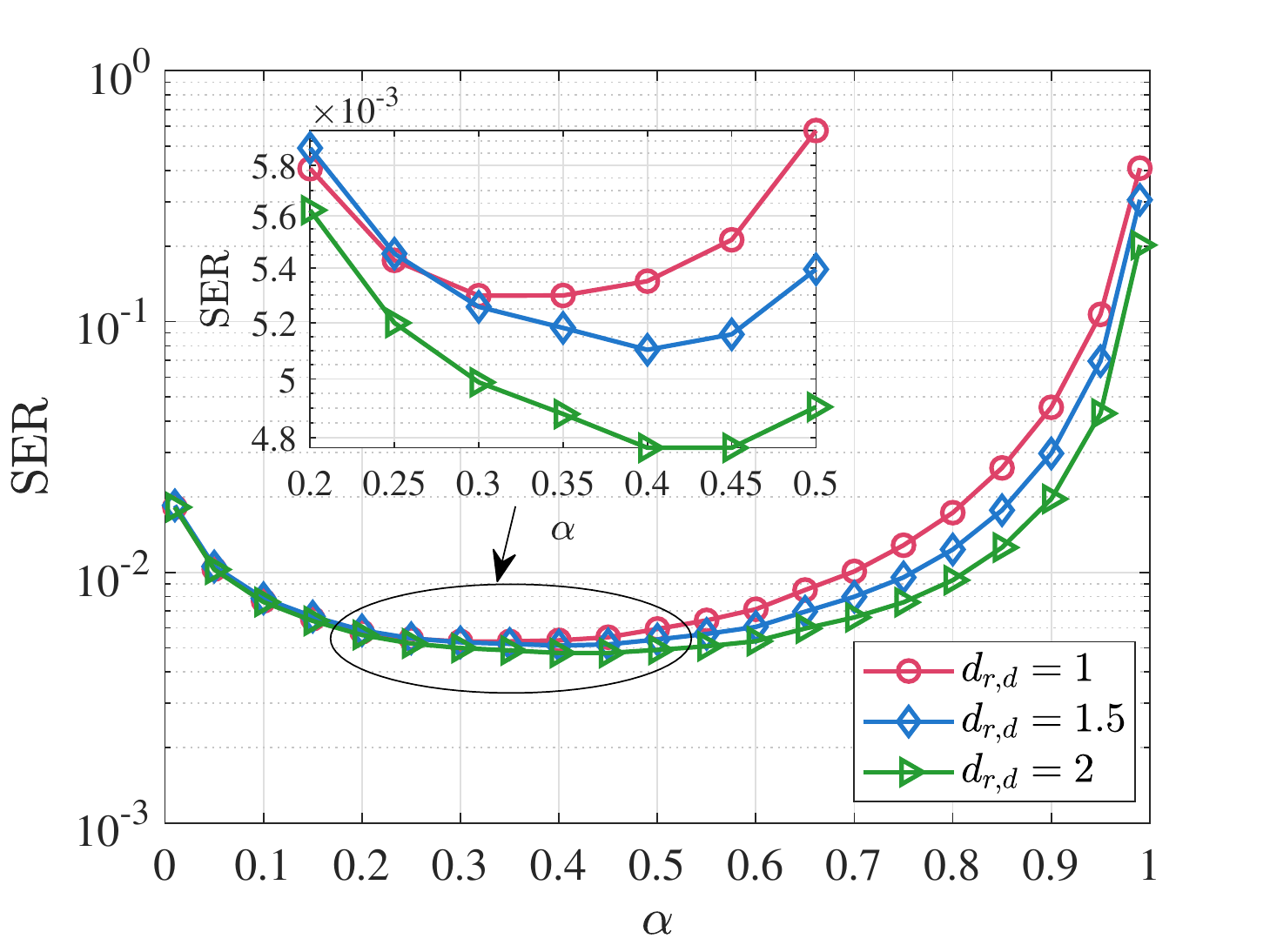}
	\caption{ \color{black}  SER performance comparison for different $R-D$ link distance  $d_{r,d}$ with respect to the TS ratio $\alpha$, where $d_{s,d}=3$,  $d_{s,r}=d_{s,d}- d_{r,d}$, and $\delta=1$, using the proposed detector with $8$-DPSK at SNR $= 40$ dB.}
	\label{fig:SER-drd-TS}
\end{figure}

{ \color{black} Fig.~\ref{fig:SER-drd-TS} compares the simulated SER for different values of the distance $d_{r,d}=1, 1.5$, and $2$ using our detector for  $d_{s,d}=3$,  $d_{s,r}=d_{s,d}- d_{r,d}$, and $\delta=1$, adopting the TS protocol. The modulation is $8$-DPSK and SNR $= 40$ dB.
It can be seen that the optimal value of $\alpha$ increases} { \color{black} as $d_{r,d}$  increases.  This is possibly because as $d_{r,d}$ increases,  the $S-R$ link becomes better while the $R-D$ link becomes worse. Then, more power is needed at $R$ for transmission (c.f. \eqref{eq:Pr-TS}) to achieve the same reliability as that of the previous.} 

\begin{figure}[t] % \usepackage{subfigure}
	\centering
	\subfigure[$M=2$, SNR = $30$ dB ]{
		\label{fig:SER-TS-M2}
		\includegraphics[width=0.6\textwidth]{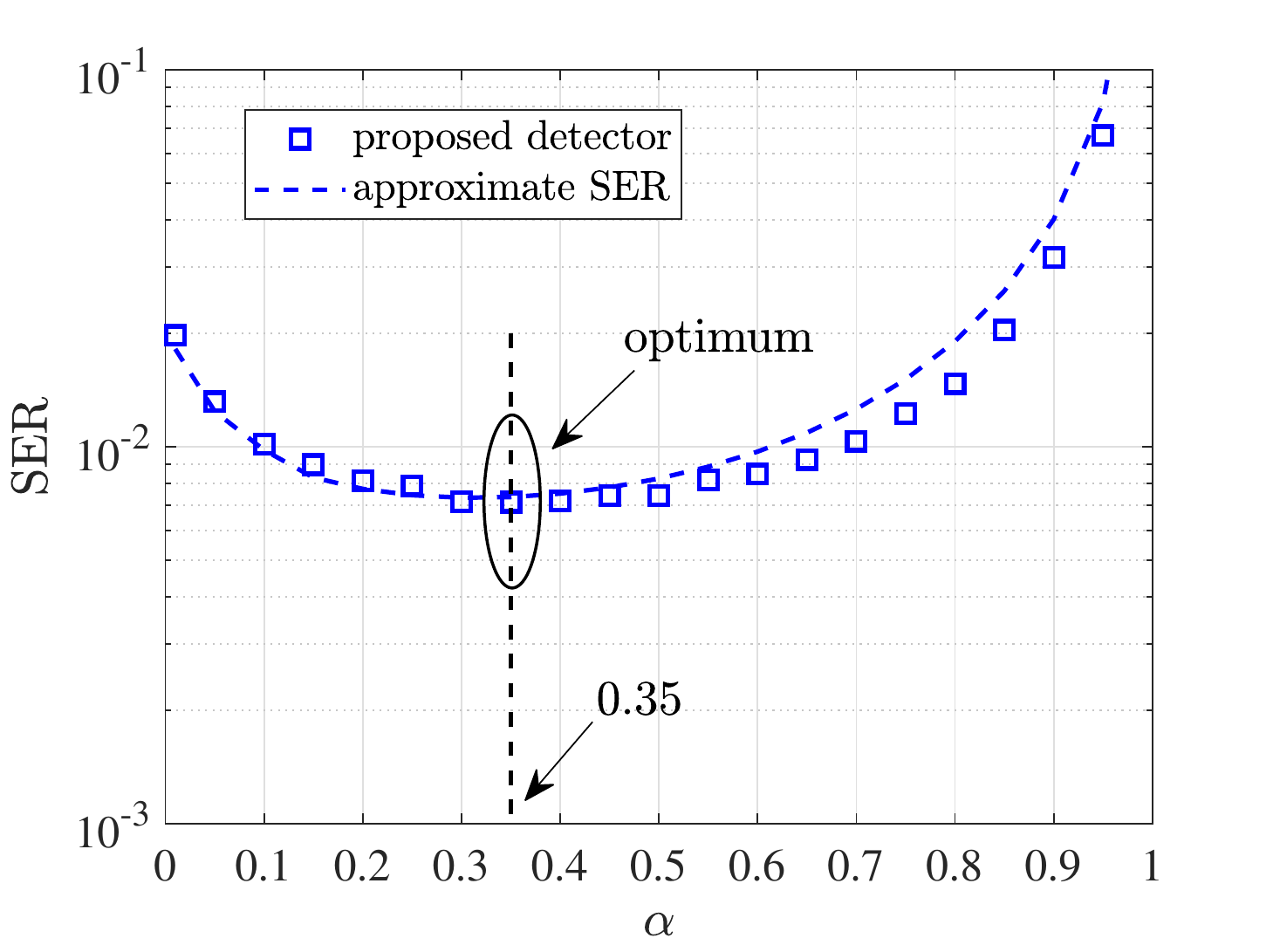}}
	%	\hspace{0.01\linewidth}
	\subfigure[$M=8$, SNR = $40$ dB]{
		\label{fig:SER-TS-M8}
		\includegraphics[width=0.6\textwidth]{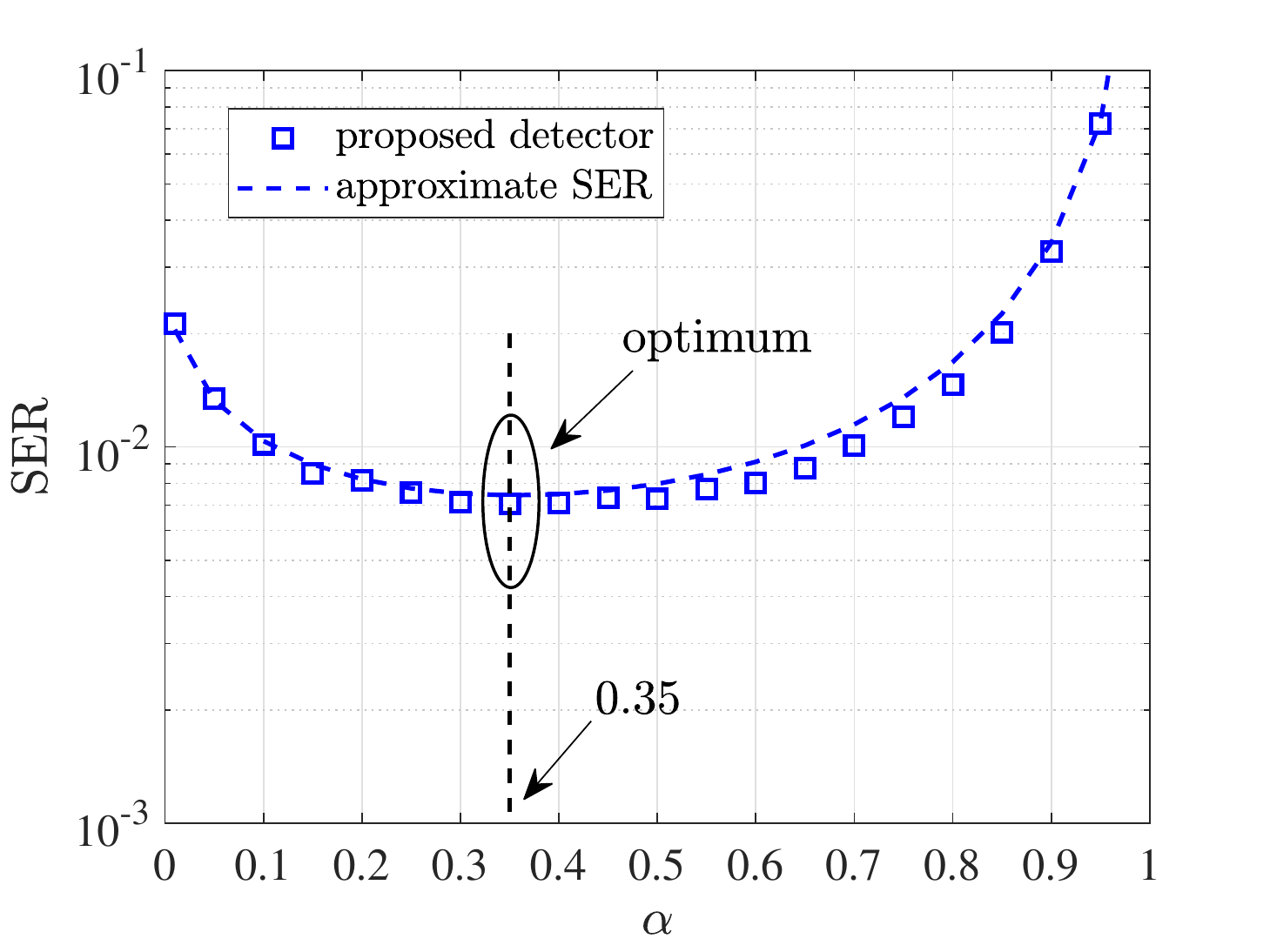}}
	\caption{\color{black}The simulated SER of the proposed detector and the proposed approximate SER expression with respect to the TS ratio $\alpha$, when $\delta=0.6$, for the SWIPT-enabled DDF relay networks with $M$-DPSK.}
	\label{fig:SER-TS-opt}
\end{figure}

{ \color{black}
Fig.~\ref{fig:SER-TS-opt} shows the simulated SER of the proposed detector and the approximate SER expression for DBPSK at SNR = $30$ dB and $8$-DPSK  at SNR = $40$ dB with respect to $\alpha$, adopting the TS protocol at the relay. We can see that the simulated SER has a unique minimum, the proposed approximate SER is quite accurate, and the minimums of the simulated and approximate SERs are consistent. The results verify that the approximate SER expression can be used to estimate the optimal TS ratio accurately. }

%%%-------------------------%%
%{ \color{black}
%	\subsection{ Detection Complexity Comparison}
%	
%}
%\end{comment}
%%---------------------------------------------------------
\clearpage
\newpage
\section{Conclusion}
\label{sec:con}
In this paper, we have proposed a near-optimal detector with linear complexity with respect to $M$ for $M$-DPSK signals, and developed new SER performance results for the SWIPT-enabled PS-based DDF single-relay network. The state-of-the-art detectors are the MLD and the approximate MLD derived in \cite{dm-df-ser-1-liu2015energy}. They serve as good performance benchmarks. However, 
their performance analysis results are not available in the literature, possibly due to the complicated non-closed-form detection metrics involving functions such as the modified Bessel function.  
Our proposed detector has a closed-form metric and its SER performance has been compared favorably with the state-of-the-art MLD and approximate MLD. We have also proposed an approximate SER expression for our detector, and this expression has been shown to be rather accurate for  various  $M$ and all values of  $\varrho$ considered, for not too small SNR values. Through analyzing this expression, we have presented the trade-off between the conditional SERs of the two scenarios where the relay detects correctly and wrongly as a function of $\varrho$. The results suggest that a good trade-off can potentially be achieved by adjusting $\varrho$. Moreover, we have proposed two methods for accurately estimating the optimal PS ratio that minimizes the SER. One is by finding the minimum of the explicit-form average approximate SER expression, which is straightforward to compute but double integral calculation is needed and is computationally expensive. The other is through finding the zero of the derived closed-form approximate derivative of the average approximate SER expression. Both methods have been verified to be quite accurate by simulation.

{\color{black} We have also extended the proposed detector for adopting the TS protocol at the relay and derived an approximate SER expression. The detector has been shown to perform very close to  the MLD and the approximate MLD for different $M$ over a range of SNRs. The SER expression has been verified to be accurate  by simulation and can be used to estimate the optimal TS ratio. 
	
In this work, we considered the linear EH model. An interesting direction is to consider the non-linear EH model \cite{EH-9069257}. Clearly, the proposed detector can still work, but since the expressions for the relay transmission power $P_r$ are different, the problems of finding the optimal PS and TS ratios need to be reformulated. Extensions to other EH protocols  at the relay (see   \cite{intro-ref6-atapattu2016optimal,concl-zr-6449245}) and to the multi-relay network are also promising directions.}
%%---------------------------------------------------------
\clearpage
\newpage
\appendix
%%---------------------------------------------------------
\subsection{Proof of \textbf{Proposition \ref{prop: 1}}}
\label{proof-prop-1}
Without loss of generality, assume $x_1$ is the source symbol and is wrongly detected to $x_v$ at $D$. Two competing symbol pairs are denoted as $(x_1, x_r)$ and $(x_v, x_u), x_v \neq x_1$.

For the scenario where the relay detects correctly, we have $x_r = x_1$. The problem of obtaining the dominating PEP terms can be formulated as
\begin{align} \label{eq:relay-correct}
\underset{x_v, x_u}{\max}  \quad & 
\left\{  \Pr [\omega_{s,d} (x_1, x_v) + \omega_{r,d} (x_1, x_v)>0 ] , \right. \notag \\
& \left.  \Pr [ \omega_{s,d} (x_1, x_v) + \omega_{r,d} (x_1, x_u) > \eta]
\right\} \notag \\
\subto \quad & (x_v,x_u) \in \mathcal{X}^2, x_v \neq x_1, x_u \neq x_v.
\end{align}
Our approach is to take all possible solutions to \eqref{eq:relay-correct} to formulate an approximate conditional SER expression $\mathcal{P}_C (\check{\gamma}) $. 

To maximize the first term in the objective, based on \eqref{eq:w-mean} and \eqref{eq:w-var}, we should maximize
\begin{align}
& \Pr [\omega_{s,d} (x_1, x_v) + \omega_{r,d} (x_1, x_v) >0 ] \notag \\
\approx  &  Q \left( \sqrt{\frac{-u_{s,d}(x_1, x_v)-u_{r,d}(x_1, x_v)}{2}}
\right) , \label{eq:case1}
\end{align}
and equivalently we should  $\minimize \quad -u_{s,d}(x_1, x_v)-u_{r,d}(x_1, x_v)$. Based on some calculations, we can obtain two possible solutions as $x_v \in \{x_2, x_M\}$. Similarly to maximize the second term, the problem is re-formulated as 
\begin{align}
\underset{x_v,x_u}{\min}	\quad  & \sqrt{-u_{s,d}(x_1, x_v)-u_{r,d}(x_1, x_u)}  + \notag \\ & \frac{\eta}{\sqrt{-u_{s,d}(x_1, x_v)-u_{r,d}(x_1, x_u)} } \notag \\
\subto \quad & (x_v,x_u) \in \mathcal{X}^2, x_v \neq x_1, x_u \neq x_v.
\end{align}

We make the assumption that $x_1$ is wrongly detected to its nearest neighbors at $D$ in this case, which is well justified when the relay detects correctly. After some calculations, the solution set is obtained as $
x_v \in \{x_2, x_M\}, x_u=x_1
$. Finally, $ \mathcal{P}_C (\check{\gamma})$ can be obtained using all dominating PEP terms. % It is obvious that the value should be half for DBPSK.  

For the scenario where the relay detects wrongly, similarly to the previous case, we take all possible solutions to \eqref{eq:relay-wrong-1} and \eqref{eq:relay-wrong-2}, respectively, to formulate an approximate conditional SER expression $\mathcal{P}_E (\check{\gamma}) $.
\begin{align}
\underset{x_v,x_u}{\max} 
\quad	& \Pr [\omega_{s,d} (x_1, x_v) + \omega_{r,d}(x_r, x_u) > 0 ]  \notag \\
\subto \quad & (x_v,x_u) \in \mathcal{X}^2, x_v \neq x_1, x_u \neq x_v.
\label{eq:relay-wrong-1} \\	
\underset{x_v}{\max}  \quad	& \Pr [\omega_{s,d} (x_1, x_v) + \omega_{r,d}(x_r, x_v) > -\eta ] \notag \\
\subto \quad & x_v \in \mathcal{X}, x_v \neq x_1.
\label{eq:relay-wrong-2}
\end{align}
After some calculations, the possible solutions are
obtained as $
x_v \in \{x_2, x_M\}, x_u=x_r, x_u \neq x_v
$, and $x_v=x_r, x_v \in \{x_2, x_M\}$, for \eqref{eq:relay-wrong-1} and \eqref{eq:relay-wrong-2}, respectively. $\mathcal{P}_E  (  \check{\gamma} )$ can be obtained accordingly.
%%-----------------------------------------------
\subsection{Proof of Monotonicity of $\tilde{\mathcal{P}}_C( \check{\gamma} )$ and $\tilde{\mathcal{P}}_E( \check{\gamma} )$}
\label{appen-mono}
Based on the expressions of $\epsilon$ and $\eta$, there is $\frac{ \partial \epsilon}{  \partial \varrho } > 0$ and $\frac{ \partial \eta}{  \partial \varrho } < 0$. For $\tilde{\mathcal{P}}_C( \check{\gamma} )$, since both $1-\epsilon$ and $Q 
\left( 
\sqrt{g_{s,d} \gamma_{s,d} +  \varrho  \delta g_{r,d}|h_{s,r}|^2 \gamma_{r,d}}
\right)$ are positive and monotonically decreasing in $\varrho$, $\tilde{\mathcal{P}}_C( \check{\gamma} )$ is monotonically decreasing in $\varrho$. By taking the derivative of $\tilde{\mathcal{P}}_E( \check{\gamma} )$ with respect to $\varrho$, we have
% \begin{small}
\begin{align}
\frac{\partial \frac{\tilde{\mathcal{P}}_E( \check{\gamma} )}{\frac{2}{M-1}}}{\partial \varrho} 
= & \frac{-\exp \left(-
	\frac{z_0^2}{2}
	\right) }{ 2 (1-\epsilon) \sqrt{2\pi g_{s,d} \gamma_{s,d}} } \frac{\partial \epsilon }{\partial \varrho} +  Q 	\left( z_0
\right) \frac{\partial \epsilon }{\partial \varrho}, \label{eq:PEa-deriv}
\end{align}
% \end{small}
where $ z_0 = \sqrt{ g_{s,d} \gamma_{s,d}} - \frac{\eta }{2   \sqrt{ g_{s,d} \gamma_{s,d}} } \overset{\eta \rightarrow \infty }{\approx} - \frac{\eta }{2   \sqrt{ g_{s,d} \gamma_{s,d}} } < 0$ . 

An accurate approximation as $Q(z_0) \approx \frac{1}{12} \exp \left(-\frac{z_0^2}{2}\right) + \frac{1}{4} \exp \left(-\frac{2}{3}z_0^2 \right), z_0 >0$ is applied to \eqref{eq:PEa-deriv}, then to prove $\frac{\partial \tilde{\mathcal{P}}_E( \check{\gamma} )}{\partial \varrho} > 0$ is equivalent to prove
% \begin{small}
\begin{align}
\frac{\frac{1}{ 2 \sqrt{ 2 \pi g_{s,d} \gamma_{s,d}} } \exp \left(-
	\frac{z_0^2}{2}
	\right) }{1-\frac{1}{12} \exp \left(-\frac{z_0^2}{2}\right) - \frac{1}{4} \exp \left(-\frac{2}{3}z_0^2 \right)} < 1-\epsilon \label{eq:mono-PE} .
\end{align}
% \end{small}
For the left side of \eqref{eq:mono-PE}, when $\eta \rightarrow \infty $, it can be approximated as $\frac{1}{ 2 \sqrt{2 \pi g_{s,d} \gamma_{s,d}} } \exp \left(-
\frac{z_0^2}{2}
\right)$, of which the value approaches $0$,
%\begin{small}
%	\begin{align}
%%		& \frac{\frac{1}{ 2 \sqrt{2 \pi g_{s,d} \gamma_{s,d}} } \exp \left(-
%%			\frac{z_0^2}{2}
%%			\right) }{1-\frac{1}{12} \exp \left(-\frac{z_0^2}{2}\right) - \frac{1}{4} \exp \left(-\frac{2}{3}z_0^2 \right)}  \notag \\
%	\approx 	& \frac{1}{ 2 \sqrt{2 \pi g_{s,d} \gamma_{s,d}} } \exp \left(-
%		\frac{z_0^2}{2}
%		\right) \rightarrow 0 \notag,
%	\end{align}
%\end{small}
while the value of the right side of \eqref{eq:mono-PE} approaches $1$. Therefore \eqref{eq:mono-PE} holds.

%%-------------------------------------------------
{\color{black}
\subsection{Derivations for Section \ref{sec:div}}
\label{app:proof-SER-ave}
\subsubsection{Derivation of \eqref{eq:SER-p2p-div}}
When $M=2$, according to \eqref{eq:SER-p2p-DM-2}, it is clear that 
\begin{align}
\lim\limits_{\bar{\gamma}_{s,r} \to \infty} \epsilon = 
\lim\limits_{\bar{\gamma}_{s,r} \to \infty}
\frac{ 1 }{2 \left[1+ \bar{\gamma}_{s,r}^{ID} (\varrho) \right]} = \frac{2-\varrho}{4(1-\varrho) T_s L_{s,r}} \frac{1}{\bar{\gamma}_{s,r}}.
\end{align}
When $M>2$, we have 
\begin{align}
\lim\limits_{\bar{\gamma}_{s,r} \to \infty} \epsilon = &
\lim\limits_{\bar{\gamma}_{s,r} \to \infty} 1.03 \sqrt{\frac{1+\cos \frac{\pi}{M}}{2 \cos \frac{\pi}{M}}}   \left[ 1-
\sqrt{\frac{ (1-\cos \frac{\pi}{M})  \frac{2(1-\varrho) T_s L_{s,r} }{2-\varrho} \bar{\gamma}_{s,r} }{1+  (1-\cos \frac{\pi}{M}) \frac{2(1-\varrho) T_s L_{s,r} }{2-\varrho} \bar{\gamma}_{s,r}  }} \
\right] \notag \\
\overset{(a)}{=} & \lim\limits_{\bar{\gamma}_{s,r} \to \infty}  \frac{1.03}{2} \sqrt{\frac{1+\cos \frac{\pi}{M}}{2 \cos \frac{\pi}{M}}}   \left[ 1-
\frac{ (1-\cos \frac{\pi}{M})  \frac{2(1-\varrho) T_s L_{s,r} }{2-\varrho} \bar{\gamma}_{s,r} }{1+  (1-\cos \frac{\pi}{M}) \frac{2(1-\varrho) T_s L_{s,r} }{2-\varrho} \bar{\gamma}_{s,r}  } \
\right] \notag \\
= & \frac{1.03}{4} \sqrt{\frac{1+\cos \frac{\pi}{M}}{2 \cos \frac{\pi}{M}}} \frac{ 2-\varrho }{(1-\cos \frac{\pi}{M}) (1-\varrho) T_s L_{s,r} } \frac{1}{\bar{\gamma}_{s,r} }  ,
\end{align}
where $(a)$ is due to 
$\lim\limits_{z \to 1} \left( 1-\sqrt{z} \right) = \frac{1}{2}(1-z)$. Then, \eqref{eq:SER-p2p-div} can be obtained.

\subsubsection{Derivation of the Average Approximate SER Expression}
	Here we denote $ \int \mathcal{P}_{C}(\check{\gamma} )
	p_{\check{\gamma}} (\check{\gamma}) d \check{\gamma} \approx \mathcal{P}_C$ and $
	\int \mathcal{P}_{E}(\check{\gamma} )
	p_{\check{\gamma}} (\check{\gamma}) d \check{\gamma} \approx \mathcal{P}_E$. 	We derive the average approximate SER for the DBPSK case ($M=2$), and the expressions for the $M>2$ case is similar. 
	
	For the second term in $\mathcal{P}_C(\check{\gamma}) $, we have 
	\begin{align}
	& \int Q 	\left( 
	\sqrt{ g_{s,d} \gamma_{s,d}} + \frac{\eta}{2}
	\frac{1}{\sqrt{  g_{s,d} \gamma_{s,d} }} 
	\right) p(\gamma_{s,d}) d \gamma_{s,d} \\
	\approx & \frac{1}{2 \bar{\gamma}_{s,d}}
	\int_0^{\infty}
	\exp \left(-
	\frac{(2 g_{s,d}\gamma +  \eta)^2}{ 8 g_{s,d}\gamma}
	\right) \exp \left(-\frac{\gamma}{\bar{\gamma}_{s,d}}\right) d \gamma \label{eq:PC-second-0} \\
	= & \frac{1}{2 \bar{\gamma}_{s,d}} 
	\frac{ \eta  \exp (-\eta/2) }{(2 g_{s,d}( g_{s,d}/2 +\bar{\gamma}_{s,d}^{-1}))^{\frac{1}{2}}}   K_{1}\left(
	\frac{\eta  \sqrt{g_{s,d}/2 +\bar{\gamma}_{s,d}^{-1}}}{ \sqrt{2g_{s,d} }} 
	\right) 
	\label{eq:PC-second-1} \\
	\approx & \frac{\sqrt{\pi}(2 g_{s,d})^{-\frac{1}{4}}}{4 \bar{\gamma}_{s,d}} \left( g_{s,d}/2+\bar{\gamma}_{s,d}^{-1} \right)^{-\frac{3}{4}} \sqrt{2 \eta } \exp (-\eta/2)   \exp \left(-
	\frac{\eta  \sqrt{ g_{s,d}/2+\bar{\gamma}_{s,d}^{-1}}}{ \sqrt{2g_{s,d} }}
	\right)\label{eq:PC-second-2} \\
	\triangleq  &  Z_1  ,
	\end{align} 
	where 
	\eqref{eq:PC-second-0} is due to $Q(x) \approx \frac{1}{2}e^{-\frac{x^2}{2}}, x >0$,  \eqref{eq:PC-second-1} is obtained according to \cite[eq. (3.471.9)]{book1-gradshteyn2014table} with $K_1(\check{\gamma})$ denoting the first-order modified Bessel function of the second kind, and  \eqref{eq:PC-second-2} is obtained  from \cite{book2-abramowitz1966handbook}. %  (therefore the approximation cannot be performed at low SNR since $\sqrt{\eta}$ induces imaginary parts)
	
	For the first term in $\mathcal{P}_C(\check{\gamma})$, by adopting $Q(x) \approx \frac{1}{2}e^{-\frac{x^2}{2}}, x >0$ similarly, we have
	\begin{align}
	& \int Q 
	\left( \sqrt{g_{s,d} \gamma_{s,d} +  \varrho  \delta g_{r,d}|h_{s,r}|^2 \gamma_{r,d}}
	\right) p(\gamma_{s,d}, \gamma_{r,d}) d \gamma_{s,d} d \gamma_{r,d} \notag \\
	\approx & \frac{1}{2 \bar{\gamma}_{s,d} \bar{\gamma}_{r,d}}
	\int_0^{\infty} \int_0^{\infty}
	\exp \left(-(g_{s,d} \gamma_{s,d} +  \varrho  \delta g_{r,d}|h_{s,r}|^2 \gamma_{r,d})/2
	\right) \exp \left(-\frac{\gamma_{s,d}}{\bar{\gamma}_{s,d}}\right)
	\exp \left(-\frac{\gamma_{r,d}}{\bar{\gamma}_{r,d}}\right)
	d \gamma_{s,d} d \gamma_{r,d} \notag \\
	= & \frac{ 2 }{ g_{s,d} \bar{\gamma}_{s,d} + 2} \frac{1}{\delta \varrho g_{r,d} |h_{s,r}|^2 \bar{\gamma}_{r,d}+2}  . \label{eq:Z-2}
	\end{align}	
	Further, by taking the expectation over  $|h_{s,r}|^2 = \frac{N_0}{P_s} \gamma_{s,r} $, we have
	\begin{align}
	& \int \frac{1}{\delta \varrho g_{r,d} |h_{s,r}|^2 \bar{\gamma}_{r,d}+2} p(\gamma_{s,r}) d \gamma_{s,r} \notag \\
	= & \frac{1}{ \bar{\gamma}_{s,r}} \int_{0}^{\infty} \frac{1}{\delta \varrho g_{r,d}  \bar{\gamma}_{r,d} \frac{N_0}{P_s} \gamma_{s,r}  +2} \exp \left(-\frac{\gamma_{s,r}}{\bar{\gamma}_{s,r}}\right) d \gamma_{s,r} \notag \\
	= & \frac{P_s}{N_0}\frac{\exp \left( \frac{2 \frac{P_s}{N_0}}{\delta \varrho g_{r,d}  \bar{\gamma}_{s,r} \bar{\gamma}_{r,d}  }\right)}{   \delta \varrho  g_{r,d} \bar{\gamma}_{s,r} \bar{\gamma}_{r,d} }   \Gamma \left(0, \frac{P_s}{N_0} \frac{2 }{\delta \varrho g_{r,d}  \bar{\gamma}_{s,r} \bar{\gamma}_{r,d} } \right),  \label{eq:int-2} 
	\end{align}
	where \eqref{eq:int-2} is obtained using \cite[eq. (3.471.13)]{book1-gradshteyn2014table}, and $\Gamma(\alpha, z) \triangleq \int_{z}^{\infty} t^{\alpha-1} \exp (-t) dt$ is the upper incomplete
	gamma function. Since 
	\begin{align}
	\Gamma(0, z) < E_1(z) \approx \exp (-z) \ln \left(
	1+\frac{1}{z}\right),
	\end{align}
	where $E_1(z)$ is the exponential integral function, we can estimate \eqref{eq:Z-2} using an upper bound as 
	\begin{align} \label{eq:Z2-up}
	Z_2 \triangleq & \frac{P_s}{N_0} \frac{ 2 \ln \left(
		1+\frac{\delta \varrho g_{r,d}  \bar{\gamma}_{s,r} \bar{\gamma}_{r,d}}{ 2 \frac{P_s}{N_0} }\right)}{ ( g_{s,d} \bar{\gamma}_{s,d} + 2)  \delta \varrho  g_{r,d}  \bar{\gamma}_{s,r} \bar{\gamma}_{r,d} }.
	\end{align}
	Based on the above, we have \eqref{eq:PC-ave}.

	For the first term in $\mathcal{P}_E(\check{\gamma})$, we have
	\begin{align}
	& \int_{0}^{\infty} Q 	\left( 
	\sqrt{ g_{s,d} \gamma_{s,d}} - \frac{\eta}{2}
	\frac{1}{\sqrt{  g_{s,d} \gamma_{s,d} }} 
	\right) p(\gamma_{s,d}) d \gamma_{s,d} \notag \\
	\approx & \frac{1}{\bar{\gamma}_{s,d}} \int_{0}^{\frac{\eta}{2 g_{s,d}  }} \exp \left(-\frac{\gamma}{\bar{\gamma}_{s,d}}\right) d \gamma   + \frac{1}{2 \bar{\gamma}_{s,d}}
	\int_{\frac{\eta }{2 g_{s,d}  }}^{\infty}
	\exp \left(-
	\frac{(2 g_{s,d}  \gamma -  \eta)^2}{ 8 g_{s,d}  \gamma}
	\right) \exp \left(-\frac{\gamma}{\bar{\gamma}_{s,d}}\right) d \gamma \label{eq:PE-first},
	\end{align}
	where 
	the first integral has a closed form solution as 
	$1-\exp \left(-\frac{\eta  }{2 g_{s,d}   \bar{\gamma}_{s,d} }
	\right)$, which approaches $0$ at high SNR. The second integral can be calculated similar to $Z_1$ as
	\begin{align}
	&  \frac{ (8 g_{s,d})^{-1/2} \eta \exp \left(\eta/2\right) }{  \left( g_{s,d}/2+\bar{\gamma}_{s,d}^{-1}\right)^{1/2}\bar{\gamma}_{s,d}} K_v \left( 
	\frac{ \eta  \sqrt{ g_{s,d}/2+\bar{\gamma}_{s,d}^{-1}}}{   \sqrt{2  g_{s,d}}}
	\right)  \notag \\
	\approx & \exp (\eta)  Z_1 \notag \\
	\triangleq & Z_3. 
	\end{align}	
	For the second term in $\mathcal{P}_E(\check{\gamma})$, we have
	\begin{align}
	& \epsilon \int_{0}^{\infty} Q \left( 
	\sqrt{  g_{s,d}   \gamma_{s,d}  } 
	\right) p(\gamma_{s,d}) d \gamma_{s,d} \approx  \frac{ \epsilon }{ g_{s,d}   \bar{\gamma}_{s,d} + 2}.
	\end{align}	
Based on the above,  we have \eqref{eq:PE-ave}.
%%---------------------------------------------------------
\subsection{Proof of \textbf{Proposition \ref{prop: 2}}}
\label{proof-prop-2}
Similarly to the PS case, for the TS case we define
\begin{align}
	\omega_{I,d} (z_1, z_2) 
	=  \Re \{ y_{I,d}^*[k] y_{I,d}[k-1] (z_2-z_1)\} /N_0 
	\sim  \mathcal{N} (u_{I,d}(z_1, z_2), W_{I,d} (z_1, z_2))
\end{align}
for $I \in \{s,r\}$, and obtain 
\begin{align}
u_{I,d}(z_1, z_2) = & T_s L_{I,d} \Re \{
x_{I}^*(z_2-z_1)
\} \frac{P_I |h_{I,d}|^2}{N_0} , \label{eq:w-mean-TS}\\
W_{I,d}(z_1, z_2) \approx  & T_s L_{I,d} |z_2-z_1|^2 \frac{P_I |h_{I,d}|^2}{N_0}, \label{eq:w-var-TS}
\end{align}
where $P_r$ is given in \eqref{eq:Pr-TS} as a function $\alpha$. 
The three problems of obtaining the dominating PEP terms can be formulated similarly as \eqref{eq:relay-correct}, \eqref{eq:relay-wrong-1}, and \eqref{eq:relay-wrong-2}. The solutions also follow the derivations in  Appendix \ref{proof-prop-1}.

}
%%---------------------------------------------
\clearpage
\newpage
\bibliographystyle{IEEEtran} 
\bibliography{ref_swipt_dm}
\ifCLASSOPTIONcaptionsoff
\newpage
\fi
%\begin{thebibliography}{00}
%\bibitem{b1} G. Eason, B. Noble, and I. N. Sneddon, ``On certain integrals of Lipschitz-Hankel type involving products of Bessel functions,'' Phil. Trans. Roy. Soc. London, vol. A247, pp. 529--551, April 1955.
%\bibitem{b2} J. Clerk Maxwell, A Treatise on Electricity and Magnetism, 3rd ed., vol. 2. Oxford: Clarendon, 1892, pp.68--73.
%\bibitem{b3} I. S. Jacobs and C. P. Bean, ``Fine particles, thin films and exchange anisotropy,'' in Magnetism, vol. III, G. T. Rado and H. Suhl, Eds. New York: Academic, 1963, pp. 271--350.
%\bibitem{b4} K. Elissa, ``Title of paper if known,'' unpublished.
%\bibitem{b5} R. Nicole, ``Title of paper with only first word capitalized,'' J. Name Stand. Abbrev., in press.
%\bibitem{b6} Y. Yorozu, M. Hirano, K. Oka, and Y. Tagawa, ``Electron spectroscopy studies on magneto-optical media and plastic substrate interface,'' IEEE Transl. J. Magn. Japan, vol. 2, pp. 740--741, August 1987 [Digests 9th Annual Conf. Magnetics Japan, p. 301, 1982].
%\bibitem{b7} M. Young, The Technical Writer's Handbook. Mill Valley, CA: University Science, 1989.
%\end{thebibliography}
\end{document}